\documentclass[pra,aps,showpacs,groupedaddress,superscriptaddress,twocolumn,nofootinbib]{revtex4-1}

\usepackage[utf8x]{inputenc}
\usepackage{color}
\usepackage{bbm} 

\usepackage{amsfonts,amsmath,amssymb,stmaryrd}

\usepackage{graphics}
\usepackage{subfigure}  % use for side-by-side figures
\usepackage{bbm} 
\usepackage{hyperref}
\usepackage{epsfig}
\usepackage{mathrsfs}
\usepackage{verbatim}
\usepackage{centernot}
\usepackage{ulem}

\renewcommand{\l}{\left(}

\renewcommand{\r}{\right)}

\newcommand{\bra}[1]{\langle#1|}

\newcommand{\ket}[1]{|#1\rangle}

\renewcommand{\H}{\hat{\mathcal{H}}}
\renewcommand{\c}{\hat{c}}
\renewcommand{\a}{\hat{a}}

\newcommand{\cd}{\hat{c}^\dagger}
\newcommand{\ad}{\hat{a}^\dagger}

\newcommand{\BZ}{\text{BZ}}

\newcommand{\hc}{\text{h.c.}}
\newcommand{\MF}{\text{MF}}

\newcommand{\ps}{\hat{\psi}}
\newcommand{\I}{\text{I}}

\newcommand{\HMF}{\mathscr{H}_\MF}
\newcommand{\HM}{\mathscr{H}}

\newcommand{\lho}{\ell_{\text{ho}}}

\usepackage{ulem}	% underline

\usepackage{array}

\usepackage{bm}	% bold in math mode
\renewcommand{\vec}[1]{\bm{#1}}

\newcommand{\ph}{\text{ph}}

\newcommand{\IB}{\text{IB}}
\newcommand{\eff}{\text{eff}}
\newcommand{\NDC}{\text{NDC}}

\begin{document}
\normalem	% changes \emph back to normal after introducing ulem package.

\title{Bloch oscillations of bosonic lattice polarons}

\author{F. Grusdt}
\affiliation{Department of Physics and Research Center OPTIMAS, University of Kaiserslautern, Germany}
\affiliation{Graduate School Materials Science in Mainz, Gottlieb-Daimler-Strasse 47, 67663 Kaiserslautern, Germany}
\affiliation{Department of Physics, Harvard University, Cambridge, Massachusetts 02138, USA}

\author{A. Shashi}
\affiliation{Department of Physics, Harvard University, Cambridge, Massachusetts 02138, USA}
\affiliation{Department of Physics and Astronomy, Rice University, Houston, Texas 77005, USA.}

\author{D. Abanin}
\affiliation{Department of Physics, Harvard University, Cambridge, Massachusetts 02138, USA}
\affiliation{Perimeter Institute for Theoretical Physics, Waterloo, Ontario N2L 6B9, Canada}
\affiliation{Institute for Quantum Computing, Waterloo, Ontario N2L 3G1, Canada}

\author{E. Demler}
\affiliation{Department of Physics, Harvard University, Cambridge, Massachusetts 02138, USA}



\date{\today}

\begin{abstract}
We consider a single impurity atom confined to an optical lattice and immersed in a homogeneous Bose-Einstein condensate (BEC). Interaction of the impurity with the phonon modes of the BEC leads to the formation of a stable quasiparticle, the polaron. We use a variational mean-field approach to study dispersion renormalization and derive equations describing non-equilibrium dynamics of polarons by projecting equations of motion into mean-field (MF) type wavefunctions. As a concrete example, we apply our method to study dynamics of impurity atoms in response to a suddenly applied force and explore the interplay of coherent Bloch oscillations and incoherent drift.  We obtain a non-linear dependence of the drift velocity on the applied force, including a sub-Ohmic dependence for small forces for dimensionality $d>1$ of the BEC. For the case of heavy impurity atoms we derive a closed analytical expression for the drift velocity. Our results show considerable differences with the commonly used phenomenological Esaki-Tsu model.
\end{abstract}

\maketitle

\section{Introduction}
\label{sec:Intro}
The problem of an impurity particle interacting with a quantum mechanical bath is one of the fundamental paradigms of modern physics. Such general class of systems, often referred to as polarons, is relevant to understanding electron properties in polar semiconductors, organic materials, doped magnetic Mott insulators and high temperature superconductors, see e.g.\cite{Alexandrov1981,PolaronsAdvMat2007,Devreese2009}. The polaron problem is closely related to the questions of macroscopic quantum tunneling \cite{caldeira1981influence,caldeira1985influence,caldeira1983path}. In the standard model of high energy physics, the way the Higgs field gives mass to various particles is also often given in terms of polaron type dressing \cite{Higgs1,Higgs2}. While the polaron problems have attracted considerable theoretical and experimental attention during the last few decades, many questions, especially addressing nonequilibrium dynamics, remain unresolved. In the present paper we study theoretically a polaron system that consists of an impurity atom confined to a species-selective optical lattice and a homogeneous BEC. The rich toolbox available in the field of ultracold atoms has already made possible a detailed experimental study of Fermi polarons \cite{Schirotzek2009,chevy2010ultra,Koschorreck2012,kohstall2012metastability,Zhang2012,Massignan_review}
and stimulated active theoretical study of both Fermi~\cite{prokof2008fermi,punk2009polaron,schmidt2011excitation,massignan2012polarons,Massignan_review} and Bose polarons~\cite{Novikov2009,Cucchietti2006,sacha2006self,kalas2006interaction,Bruderer2008,Bruderer2008a,Bruderer2007,Privitera2010,Casteels2011a,Casteels2012,Casteels2011,Tempere2009,blinova2013single,Rath2013,Shashi2014RF}. First experiments have also started to explore physics connected to the Bose polaron~\cite{Catani2012,schmid2010dynamics,Spethmann2012,Fukuhara2013,Scelle2013}. Additionally, cold atomic ensembles are well suited to the investigation of non-equilibrium phenomena~\cite{Greiner2002,kinoshita2006quantum,hofferberth2007non,haller2009realization,cheneau2012light} since they are very well isolated from the environment and their parameters can be tuned dynamically. There is thus a growing interest in out of equilibrium polaron problems~\cite{Astrakharchik2004,Bruderer2008,Bruderer2010,Johnson2011,Schecter2012a,Knap2014,Fukuhara2013,Dasenbrook2013,Scelle2013,Fukuhara2013} which remained out of reach in solid-state systems due to short equilibration times.

 %%%%%%%%%%%%%%%%%%%%%%%%%%%%%%%%%%%%%%%%%%%%%%%%%%%%%
\begin{figure}[h!]
\centering
\includegraphics[width=0.5\textwidth]{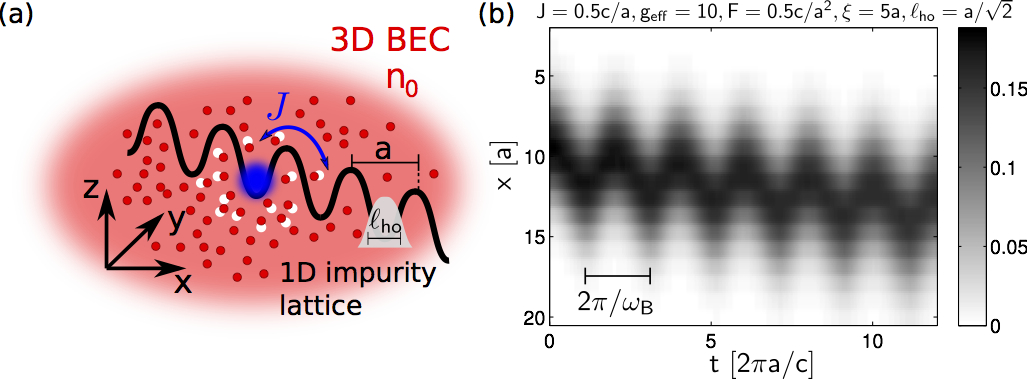}
\caption{(Color online) (a) An impurity (blue) is immersed in a homogeneous 3D BEC (red) and constrained to the lowest band of a 1D optical lattice. Strong interactions with the Bose gas lead to polaron formation and a modified dispersion.
 (b) Applying a constant force to the impurity alone results in polaron Bloch oscillations (BO). Although the speed of sound $c$ is never exceeded, BO are superimposed by a constant drift velocity $v_\text{d}$ as well as diffusion of the polaron wavepacket.}
\label{fig:setup}
\end{figure}
%%%%%%%%%%%%%%%%%%%%%%%%%%%%%%%%%%%%%%%%%%%%%%%%%%%%%

We consider the system shown in Fig.\ref{fig:setup}(a) consisting of a single impurity atom, confined to the lowest Bloch band of an optical lattice (hopping $J$, lattice constant $a$), immersed in a weakly interacting Bose-Einstein condensate (BEC). The BEC hosts gapless Bogoliubov phonons which can scatter off the impurity, leading to polaron formation \cite{Novikov2009,Casteels2011a,Casteels2012,Casteels2011,Tempere2009,Rath2013,Shashi2014RF}. We subject the impurity to a constant force and examine in detail how the dynamics of the impurity will be affected by its interaction with the surrounding phonons. This is the central focus of the present article.

While it is well known that an isolated quantum mechanical particle in a lattice will undergo coherent Bloch oscillations (BO) when subject to a constant force, it is less obvious that a composite quasiparticle, i.e. an impurity coupled to a phonon bath, will display coherent BO. Here we establish that the Bose polaron can indeed undergo BO. Next, by calculating the renormalized shape of the dispersion relation of the polaron, we show that phonon dressing has a pronounced effect on BO, which can be observed in experiments by measuring the real time dynamics of impurity atoms. Such experiments can be done using recently developed quantum gas microscopes with single atom resolution~\cite{Gericke2008,Bakr2009,Sherson2010}, but it should be noted that single-site resolution in the optical lattice is not a necessary requirement to observe polaron BO.

Polarons in optical lattices were considered earlier by Bruderer et al. \cite{Bruderer2007,Bruderer2008}, however, in contrast to our work, they considered the strong coupling limit of the so-called ``small'' polaron where impurity hopping is sub-dominant to phonon coupling. This regime was further studied in \cite{Privitera2010}, and is indeed the traditional approach to lattice polarons \cite{holstein1959studies,alexandrov2009advances} in solid state systems. We use an alternative approach which is flexible enough to describe both limits of weakly coupled ``large'' polarons and heavy impurities, which can both be achieved in experiments with cold atomic Bose-Fermi~\cite{truscott2001observation,Stan2004,gunter2006bose,Inouye2004,fukuhara2009all,Wu2012a} or Bose-Bose~\cite{Catani2008,Shin2008,Pilch2009,wernsdorfer2010lattice,mccarron2011dual,Spethmann2012} mixtures. Our approach is based on a variational mean-field (MF) ansatz~\cite{Alexandrov1995,BeiBing2009}, which we generalized earlier to study spectral properties of unconfined impurities in Bose gases~\cite{Shashi2014RF}. By extending this approach to lattice impurities, we calculate the full non-equilibrium dynamics of polaron BO. In particular, we find that polaron formation takes place on a timescale $\xi/c$ set by the BEC ($\xi$ is the healing length, $c$ the speed of sound), and subsequently we observe pronounced BO, see FIG.\ref{fig:setup} (b). Additionally, to gain insight into the nature of the coherent polaron dynamics, we introduce an analytic ``adiabatic approximation'' which correctly predicts the predominant characteristics of the polaron trajectory in the subsonic regime, e.g. overall shape, and frequency of oscillations.  As an added advantage of our approach, in contrast to the strong coupling polaron approximation, we can approach the supersonic regime, near which we find strong decoherence of the BO in connection with a large drift velocity $v_\text{d}$, i.e. a net polaron current. Such incoherent transport can only be sustained in the presence of decoherence mechanisms, and indeed we observe phonon emission in this regime.

Historically, the study of the interplay between coherent BO and inelastic scattering (e.g. on phonons) was pioneered in the solid state context by Esaki and Tsu \cite{Esaki1970}, who derived a phenomenological relation between the driving force $F$ and the net (incoherent) current $v_\text{d}$, and proposed a generic Ohmic regime for weak driving, i.e. $v_\text{d} \sim F$. The precision of ultra-cold atom experiments allowed a detailed verification of the Esaki-Tsu model in thermal gases \cite{Ott2004}, and thus triggered theoretical interest in this topic \cite{Ponomarev2006,Kolovsky2008}. While all these works focused on non-condensed gases, Bruderer et al. \cite{Bruderer2008} considered a 1D BEC where the phonons provide an Ohmic bath (see e.g. \cite{DallaTorre2010}) and established a finite current (i.e. $ v_\text{d} \neq 0$) even for subsonic impurities, with a current-force relation $v_\text{d}(F)$ of a shape similar to that predicted by Esaki and Tsu.

In this article we address the question how polaron BO decohere, and in particular how the polaron drift velocity depends on the driving force, for condensates in arbitrary dimensions $d=1,2,3,...$. In the weak driving regime, we find that the drift current strongly depends on dimensionality $d$ and deviates from the Ohmic behavior predicted by the phenomenological Esaki-Tsu relation. We show that a quantitative description of polaron drift can be obtained by applying Fermi's golden rule to calculate the phonon emission of oscillating impurity atoms. This analysis correctly reproduces the current-force relation $v_\text{d} \sim F^d$ observed in our numerics for weak driving.
 
The paper is organized as follows. In Sec.\ref{sec:Model} we introduce our model and employ the Lee-Low-Pines unitary transformation to make use of the discrete translational invariance (by a lattice period) of our problem. Then in Sec.\ref{sec:StaticProps} we discuss the ground state of the impurity-Bose system in the presence of a lattice, and calculate the renormalized polaron dispersion. We also present the MF phase diagram which shows where the subsonic to supersonic transition takes place. In Sec.\ref{sec:adiabBO} we discuss polaron BO within the adiabatic approximation. We also show that direct imaging of real-space impurity trajectories reveals the renormalized polaron dispersion. How non-adiabatic corrections modify BO is studied in Sec.\ref{sec:Dynamics} using a time-dependent variational wavefunction. In Sec.\ref{sec:NonAdiabatic} we discuss incoherent polaron transport and present numerical as well as analytical results for its dependence on the driving force. Finally in Sec.\ref{sec:Summary} we summarize our results.

\section{The model}
\label{sec:Model}

In this section we present our theoretical model, starting from the microscopic Hamiltonian in Subsection \ref{subsec:microModel}. We subsequently simplify the latter by applying Bogoliubov theory for the BEC as well as nearest-neighbor tight-binding approximation for the free impurity. Then we derive the corresponding impurity-boson interaction, which requires careful treatment of the two-particle scattering problem in order to derive correct system parameters. Having established the connection to microscopic properties, we discuss realistic numbers and introduce a dimensionless polaron coupling constant. In the second part \ref{sec:LLP} we apply the Lee-Low-Pines transformation to our model, which is at the heart of our formalism and makes conservation of the polaron quasimomentum explicit.  

Here, as well as in the subsequent three sections, we will focus on the case of a three dimensional BEC ($d=3$), but an analogous analysis can be done for dimensions $d=1,2$. We will discuss the difference in dynamics of systems of different dimensionality in Section \ref{sec:NonAdiabatic}.

\subsection{Microscopic origin}
\label{subsec:microModel}
We start by considering weakly interacting bosons of mass $m_\text{B}$ in three spatial dimensions ($d=3$) and at zero temperature, which will be described by the field operator $\hat{\phi}(\vec{r})$. Next we introduce a single impurity of mass $m_\text{I}$, which can be described by a second field operator $\ps(\vec{r})$. The impurity is furthermore confined to a deep species-selective optical lattice which is completely immersed in the surrounding Bose gas. For concreteness we assume the lattice to be one-dimensional (pointing along $\vec{e}_x$), but our analysis can easily be carried over to arbitrary lattice dimensions. The bosons and impurity interact via a contact interaction of strength $g_\text{IB}$. Since we wish to study transport properties of the dressed impurity, we will also consider a constant force $F$ acting on the impurity alone. In experiments this force can e.g. be applied using a magnetic field gradient\cite{Anderson1998,Palzer2009,Atala2012}. The microscopic Hamiltonian of this system reads ($\hbar=1$)
\begin{multline}
 \H = \int d^3\vec{r} \left\{ \hat{\phi}^{\dagger}(\vec{r}) \left[ -\frac{\nabla^2}{2 m_\text{B}} + \frac{g_{\text{BB}}}{2} \hat{\phi}^\dagger(\vec{r}) \hat{\phi}(\vec{r})\right] \hat{\phi}(\vec{r}) 
 \right. \\ \left.  + \hat{\psi}^{\dagger}(\vec{r}) \left[ -\frac{\nabla^2}{2 m_\text{I}} + V_\text{I}(\vec{r}) + g_{\IB} \hat{\phi}^{\dagger}(\vec{r}) \hat{\phi}(\vec{r}) \right] \hat{\psi}(\vec{r}) \right\}.
 \label{eq:microHam}
\end{multline}
Here $V_\I(\vec{r})$ denotes the optical lattice potential seen by the impurity and $g_{\text{BB}}$ is the boson-boson interaction strength. The impurity is confined to a spatial region $y^2 + z^2 \lesssim \l \lambda / 2 \r^2 $, where the optical potential is assumed to have a form
\begin{equation}
V_{\I}(\vec{r}) = V_0 \left[ \sin^2 \l k_0 x \r + \sin^2 \l k_0 y \r + \sin^2 \l k_0 z \r \right] - F x ,
\end{equation}
including a linear potential $- F x$ describing the constant force acting on the impurity. Here $k_0=2 \pi / \lambda$ is the optical wave vector used to create the lattice potential.

\subsubsection{Free Hamiltonians}
We will assume the optical lattice to be sufficiently deep to employ nearest-neighbor tight-binding approximation. The operator 
$\cd_j$ (written in second quantization) creates a particle at site $j$. The corresponding Wannier functions can be approximated by local oscillator wavefunctions,
\begin{equation}
 w_j(\vec{r}) = \l \pi \lho^2 \r^{-3/4} e^{- \l \vec{r} - j a \vec{e}_x \r^2 / (2 \lho^2)},
 \label{eq:WannierFct}
\end{equation}
where $\lho=1/\sqrt{m_\text{I} \omega_0}$ is the oscillator length in a micro trap and $\omega_0=2 \sqrt{V_0 E_{\text{r}}}$ the corresponding micro trap frequency, given by the recoil energy $E_\text{r}=k_0^2/2 m_\text{I}$ \cite{Bloch2008}. This gives rise to an effective hopping $J$ between lattice sites, such that -- after inclusion of the uniform force -- the free impurity Hamiltonian reads
\begin{equation}
\H_{\text{I}} =  - J \sum_j \l \cd_{j+1} \c_j + \hc \r - F \sum_j j a ~ \cd_j \c_j. 
\end{equation}

In the absence of the impurity, bosons condense and form a BEC. In the spirit of Refs.~\cite{Bruderer2007,Bruderer2008,Tempere2009}, we will assume that the impurity-boson interaction does not significantly alter the many-body spectrum of the bath, allowing us to treat the bosons as an unperturbed homogeneous condensate within the Bogoliubov approximation~\cite{Pitaevskii2003}. Consequently, the BEC is fully characterized by the speed of sound $c$, the healing length $\xi$ and its density $n_0$. The elementary excitations of the system are gapless (Bogoliubov) phonons $\a_{\vec{k}}$, the dispersion relation of which reads
\begin{equation}
\omega_{\vec{k}} = c k \sqrt{ 1+\frac{1}{2} \xi^2 k^2 }.
\end{equation}
Here $\vec{k} \in \mathbb{R}^3$ is the 3D phonon momentum (with $k$ denoting its absolute value), and the free boson Hamiltonian is given by
\begin{equation}
\H_\text{B} = \int d^3 \vec{k} ~ \omega_k \ad_{\vec{k}} \a_{\vec{k}}.
\end{equation}
In this paper $\int d^3 \vec{k} = \int_{-\infty}^{\infty} dk_x dk_y dk_z$ denotes the integral over all momenta from the entire $\vec{k}$-space.

\subsubsection{Impurity-Boson interaction}
In the discussion of the interaction Hamiltonian describing impurity-boson scattering, we restrict ourselves to the tight binding-limit. This allows us to expand the impurity field in terms of Wannier orbitals,
\begin{equation}
\ps(\vec{r}) = \sum_j \c_j w_j(\vec{r}).
\end{equation}
Using this decomposition, Eq.\eqref{eq:microHam} yields the following expression for the impurity-boson Hamiltonian,
\begin{equation}
\H_\IB = g_\IB \sum_j \cd_j \c_j \int d^3 \vec{r} ~ |w_j(\vec{r})|^2 \hat{\phi}^\dagger(\vec{r}) \hat{\phi}(\vec{r}),
\label{eq:HIBtwoBody}
\end{equation}
where we neglected phonon-induced hoppings (the validity of this approximation will be discussed further below). 

An important question is how the interaction strength $g_\IB$ in the simplified model \eqref{eq:HIBtwoBody} above relates to the measurable impurity-boson scattering length $a_\IB$. While for unconfined impurities this relation is usually derived from the Lippmann-Schwinger equation describing two-particle scattering, it is more involved for an impurity confined to a lattice. In this case the new scattering length $a_\IB^\eff$ has to be distinguished from its free-space counter part, and can even be substantially modified due to lattice effects \cite{Massignan2006,Nishida2010}. Furthermore, also the effective range $r_\IB^\eff$ of the interaction between a free boson and an impurity confined to a lattice can be modified by the lattice. We can take this effect into account in our model by choosing a proper extent $\lho$ of Wannier functions in Eq.\eqref{eq:HIBtwoBody}.

In the follwing we will not calculate the numerical relation between $a_\IB^\eff$ (or $r_\IB^\eff$) in the lattice and its free-space counterpart $a_\IB$. Instead we assume that these numbers are known -- either from numerical calculations \cite{Massignan2006,Nishida2010} or from an experiment \cite{Lamporesi2010} -- and work with the effective model Eq.\eqref{eq:HIBtwoBody}. In the Appendix \ref{appdx:InteractionStrengthAndScatteringLength} we discuss in detail how $a_\IB^\eff$ and $r_\IB^\eff$ relate to our model parameters $g_\IB$ and $\lho$, in the tight-binding case.

\subsubsection{Polaron Hamiltonian}
\label{subsubsec:PolaronHamiltonian}
Next, in order to derive a simplified Hamiltonian, we replace bare bosons $\hat{\phi}(\vec{r})$ by Bogoliubov phonons $\a_{\vec{k}}$. In doing so, we will assume sufficiently weak interactions between the impurity and the bosons, thus causing negligible quantum depletion of the condensate. This allows us to neglect two-phonon processes corresponding to terms like $\a_{\vec{k}} \a_{\vec{k}'}$ in the full Hamiltonian. As shown in \cite{Bruderer2007} and -- via a different approach --  in the Appendix \ref{sec:EffecHam} of this paper, it is justified for 
\begin{equation}
|g_\IB| \xi^{-3} \ll 4 c / \xi.
\end{equation} 

Under this condition, and provided that phonon-induced hopping can be neglected, we arrive at a Hamiltonian which is closely related to the one derived by Fr\"ohlich \cite{Froehlich1954},
\begin{multline}
 \H = \int d^3 \vec{k} \Bigl\{ \omega_k \ad_{\vec{k}} \a_{\vec{k}} + \sum_j \cd_j \c_j e^{i k_x a j} \l \ad_{\vec{k}} + \a_{-\vec{k}} \r V_{k} \Bigr\} \\
 + g_\IB n_0 - J \sum_j \l \cd_{j+1} \c_j + \hc \r - F \sum_j j a ~ \cd_j \c_j,
  \label{eq:Hfund}
\end{multline}
as we show in a more detailed calculation in Appendix \ref{sec:EffecHam}. Here the phonon-impurity interaction is characterized by
\begin{equation}
 V_k = (2 \pi)^{-3/2} \sqrt{n_0} g_\IB \l \frac{(\xi k)^2}{2 + (\xi k)^2} \r^{1/4} e^{-k^2 \lho^2/4},
\end{equation}
where $\lho$ is the oscillator length in the tight-binding Wannier function, see Eq.\eqref{eq:WannierFct}. The second term in the first line of Eq.\eqref{eq:Hfund} $\sim \cd_j \c_j$ describes scattering of phonons on an impurity localized at site $j$ (with amplitude $V_k$). This term thus breaks the conservation of total phonon momentum (and number), and we stress that phonon momenta $\vec{k}$ can take arbitrary values $\in \mathbb{R}^3$, not restricted to the Brillouin zone (BZ) defined by the impurity lattice \footnote{When the host BEC atoms are subject to a lattice potential, the phonon momenta $\vec{k}$ appearing in Eq \eqref{eq:Hfund} should be restricted to the BZ. In this paper we consider only the case when BEC atoms are \emph{not} affected by the optical lattice.}.

Phonon induced tunneling, which in the nearest neighbor case has the form
\begin{equation}
\H_{J-\ph} = \sum_j \cd_{j+1} \c_j e^{i k_x a j} \l \ad_{\vec{k}} + \a_{-\vec{k}} \r V_{k}^{(1)} + \hc,
\end{equation}
can be neglected when $|V_k^{(1)}| \ll | V_k|$. Using the result for $V_k^{(1)}$ from Eq.\eqref{eq:Vkn} in Appendix \ref{sec:EffecHam}, this condition reads in terms of Wannier functions
\begin{equation}
 |\bra{w_{j+1}} e^{i \vec{k} \cdot \vec{r}} \ket{w_j}| \ll  |\bra{w_j} e^{i \vec{k} \cdot \vec{r}} \ket{w_j}|.
\end{equation}
It is automatically fulfilled for a sufficiently deep lattice, or provided that $ k a \ll 1$ for typical phonon momenta $k$. In the latter case we may expand $e^{i \vec{k} \cdot \vec{r}} \approx 1 + i  \vec{k} \cdot \vec{r}$ in the overlap above. The zeroth order term thus vanishes because of orthogonality of Wannier functions, and the leading order term is $|\bra{w_{j+1}} \vec{k} \cdot \vec{r} \ket{w_j}| \lesssim a k \ll 1$.

\subsubsection{Coupling constant and relation to experiments}
As we discussed earlier, in contrast to Refs.~\cite{Bruderer2007,Bruderer2008} we want our analysis to be applicable to the case of  "large'' polarons, characterized by a phonon cloud with radius exceeding the impurity lattice spacing, $\xi > a$. Such polarons are typical when interactions are weak compared to impurity hopping, leading to a loosely confined phonon cloud. Indeed, it is convenient to measure the strength of interactions by defining the following dimensionless coupling constant,
\begin{equation}
g_\eff = \sqrt{\frac{n_0 g_{\text{IB}}^2}{\xi c^2}},
\end{equation}
which appears naturally in our formalism. It describes the ratio between characteristic impurity-boson interactions $E_\IB = g_\IB \sqrt{n_0 \xi^{-3}}$ and typical phonon energies $E_\text{ph} = c/\xi$, $g_\eff = E_\IB / E_\text{ph}$. Let us note that Tempere et al. \cite{Tempere2009} introduced an alternative dimensionless coupling constant $\alpha= \frac{2}{\pi} m_\text{red}^{-2} n_0 g_\IB^2$, where $m_\text{red} =1 / \l 1/ m_{\text{I}} + 1 / m_\text{B}\r$ is the reduced mass. It is related to our choice by
\begin{equation}
\alpha = \frac{1}{\pi} \left[1 + \frac{m_\text{B}}{m_\text{I}} \right]^{-2} g_\eff^2.
\end{equation}
Because in this expression the impurity mass enters as an additional parameter, which is not required to calculate $g_\eff$, we prefer to use $g_\eff$ instead of $\alpha$ in this work.

For experimentally realized Bose-Bose \cite{Catani2008,Pilch2009,Spethmann2012} or Bose-Fermi mixtures \cite{Shin2008,Wu2012a} we find that background interaction strengths are of the order $g_\eff \sim 1$, but using Feshbach resonances values as large as $g_\eff = 15$~\cite{Shashi2014RF} should be within reach. For standard Rubidium BECs characteristic parameters are $\xi \approx 1 \mu \text{m}$, $c\approx 1\text{mm}/\text{s}$ and for Rubidium in optical lattices one typically has hoppings $J \lesssim 1 \text{kHz}$ \cite{Bloch2008}.

\subsection{Lee-Low-Pines transformation}
\label{sec:LLP}
To make further progress, we will now simplify the Hamiltonian \eqref{eq:Hfund}. To this end we make use of the Lee-Low-Pines transformation, making conservation of polaron quasimomentum explicit, and include the effect of the constant force $F$ acting on the impurity. To do so, we apply a time-dependent unitary transformation,
\begin{equation}
\hat{U}_\text{B}(t) = \exp \l i \omega_{\text{B}} t \sum_j j \cd_j \c_j \r,
\label{eq:BOtransform}
\end{equation}
where $\omega_\text{B} = a F$ denotes the BO frequency of the bare impurity. In the new basis the (time-dependent) Hamiltonian reads
\begin{equation}
\tilde{\mathcal{H}}(t) = \hat{U}_\text{B}^\dagger(t)  \H \hat{U}_\text{B}(t) - i \hat{U}_\text{B}^\dagger(t) \partial_t \hat{U}_\text{B}(t),
\label{eq:Hefft}
\end{equation}
and we introduce the quasimomentum basis in the usual way,
\begin{equation}
 \c_q := \l L /a \r^{-1/2} \sum_j e^{i q a j} \c_j,
  \label{eq:BOoscillatingOps}
\end{equation}
where $L$ denotes the total length of the impurity lattice and $q = -\pi/a,...,\pi/a$ is the impurity quasimomentum in the BZ. The transformation \eqref{eq:BOtransform} allows us to assume periodic boundary conditions for the Hamiltonian \eqref{eq:Hefft}, despite the presence of a constant force $F$.

In a second step, we apply the Lee-Low-Pines unitary transformation \cite{Lee1953}, described by
\begin{equation}
 \hat{U}_{\text{LLP}}=e^{i \hat{S}}, \quad \hat{S}= \int d^3\vec{k} ~ k_x \ad_{\vec{k}} \a_{\vec{k}}  ~ \sum_{j}  a  j  \cd_j \c_j.
 \label{eq:polaronTrofo}
\end{equation}
The new frame, obtained by applying the transformation $\hat{U}_{\text{LLP}}$ to our system, will be called \emph{polaron frame} in the following. Here $k_x=\vec{k} \cdot \vec{e}_x$ denotes the $x$-component of $\vec{k}$ \footnote{In practice, when doing calculations, we find it convenient to introduce spherical coordinates around the $x$-axis, such that $k_x = k \cos \vartheta$ with $\vartheta$ the polar angle. In these coordinates rotational symmetry around the direction of the impurity lattice $\vec{e}_x$ is made explicit, and all expressions are independent of the azimuthal angle $\varphi$.}. 
The action of the Lee-Low-Pines transformation on an impurity can be understood by noting that it can be interpreted as a displacement in quasimomentum space. Such a displacement  $q \rightarrow q + \delta q$ (modulo reciprocal lattice vectors $2 \pi /a$) is generated by the unitary transformation $e^{i \delta q \hat{X}}$, where the impurity position operator is defined by $\hat{X} = \sum_j a j \cd_j \c_j$. Comparing this to Eq.\eqref{eq:polaronTrofo} yields $\delta q = \int d^3 \vec{k}~ k_x \ad_{\vec{k}} \a_{\vec{k}}$, which is the total phonon momentum operator. Thus we obtain
\begin{equation}
 \hat{U}_{\text{LLP}}^\dagger \c_q  \hat{U}_{\text{LLP}} =  \c_{q+\delta q}.
 \label{eq:ULLP1}
\end{equation}
For phonon operators, on the other hand, transformation \eqref{eq:polaronTrofo} corresponds to translations in real space by the impurity position $\hat{X}$ and one can easily see that
\begin{equation}
\hat{U}_{\text{LLP}}^\dagger \a_{\vec{k}} \hat{U}_{\text{LLP}} = e^{i \hat{X} k_x} \a_{\vec{k}}.
\label{eq:ULLP2}
\end{equation}

Now we apply the Lee-Low-Pines transformation to the Hamiltonian Eq.\eqref{eq:Hfund}. To this end, we first write the free impurity Hamiltonian in quasimomentum space, 
\begin{equation}
\H_\text{I} = -2 J \sum_{q \in \BZ} \cd_q \c_q \cos (a q).
\end{equation}
Next we make use of the fact that only a single impurity is considered, i.e. $\sum_{q\in \BZ} \cd_q \c_q =1$, allowing us to simplify
\begin{equation}
\cd_j \c_j e^{i k_x  \hat{X}  } = \cd_j \c_j e^{i k_x  a j} .
\label{eq:smplifyeikhatX}
\end{equation}
Note that although the operator $\hat{X}$ in Eq.\eqref{eq:smplifyeikhatX} consists of a summation over all sites of the lattice, in the case of a single impurity the prefactor $\cd_j \c_j$ selects the contribution from site $j$ only. 

We proceed by employing Eqs.\eqref{eq:ULLP1} - \eqref{eq:smplifyeikhatX} and arrive at the Hamiltonian
\begin{widetext}
\begin{equation}
\H(t) = \hat{U}_{\text{LLP}}^\dagger \tilde{\mathcal{H}}  \hat{U}_{\text{LLP}} = \sum_{q \in \BZ} \cd_{q} \c_q \left\{ \int d^3 \vec{k} \Bigl[ \omega_k \ad_{\vec{k}} \a_{\vec{k}} + V_k \l \ad_{\vec{k}} + \a_{\vec{k}} \r \Bigr]  - 2 J \cos \l a q - \omega_\text{B} t - a \int d^3 \vec{k}' ~k_x' \ad_{\vec{k}'} \a_{\vec{k}'} \r + g_\IB n_0 \right\}.
 \label{eq:polaronHam}
\end{equation}
\end{widetext}
Let us stress again that this result is true only for a single impurity, i.e. when $\sum_{q \in \BZ} \cd_q \c_q=1$. We find it convenient to make use of this identity and pull out $\sum_{q \in \BZ} \cd_q \c_q$ everywhere to emphasize that the Hamiltonian factorizes into a part involving only impurity operators and a part involving only phonon operators. Notably the Hamiltonian \eqref{eq:polaronHam} is time-dependent and non-linear in the phonon operators. From the equation we can moreover see that, in the absence of a driving force $F=0$ (corresponding to $\omega_{B} = 0$), the total \emph{quasi}momentum $q$ in the BZ is a conserved quantity. We stress, however, that the total \emph{phonon}-momentum $\int d^3 \vec{k} ~ \vec{k} \ad_{\vec{k}} \a_{\vec{k}}$ of the system is not conserved.

Even in the presence of a non-vanishing force $F\neq 0$ the Hamiltonian is still block-diagonal for all times, 
\begin{equation}
\H(t) = \sum_{q \in \BZ} \cd_q \c_q \H_q(t),
\label{eq:qtConserved}
\end{equation}
and quasimomentum evolves in time according to 
\begin{equation}
q(t) = q - F t,
\label{eq:qt}
\end{equation}
i.e. $\H_q(t) = \H_{q(t)}(0)$. This relation has the following physical meaning: if we start with an initial state that has a well defined quasimomentum $q_0$, then the quasimomentum of the system remains a well defined quantity. The rate of change of the quasimomentum is given by $F$, i.e. $q(t) = q_0 - F t$. Thus states that correspond to different initial momenta do not mix in the time-evolution of the system.

\section{Polarons without the drive: dispersion renormalization}
\label{sec:StaticProps}
Before turning to the nonequilibrium problem of polaron BO in the next section, we discuss the equilibrium properties at $F=0$. Because we employed the Lee-Low-Pines canonical transformation above, quasimomentum $q$ is explicitly conserved in the Hamiltonian \eqref{eq:polaronHam}. This enables us to treat every sector of fixed $q$ separately for the characterization of the equilibrium state.

We begin the section by introducing the MF polaron wavefunction in Subsection \ref{subsec:polaronGSmodel}, where we also minimize its variational energy. This readily gives us the renormalized polaron dispersion, the properties of which we discuss in Subsection \ref{subsec:polaronGSresults}. There we moreover present the MF polaron phase diagram.

\subsection{Model and MF ansatz}
\label{subsec:polaronGSmodel}
To describe the polaron ground state we apply the variational ansatz of uncorrelated coherent phonon states, which has been used successfully for polarons in the absence of a lattice \cite{Alexandrov1981,BeiBing2009,Shashi2014RF},
\begin{equation}
\ket{\Psi_q^\MF} = \prod_{\vec{k}} \ket{\alpha_{\vec{k}}^\MF}.
\label{eq:MFansatz}
\end{equation}
Here $\ket{\alpha_{\vec{k}}^\MF}$ denotes coherent states with amplitude $\alpha_{\vec{k}}^\MF \in \mathbb{C}$,
\begin{equation}
\ket{\alpha_{\vec{k}}^\MF} = \exp \left[ \alpha_{\vec{k}}^\MF \ad_{\vec{k}} - \l \alpha_{\vec{k}}^{\MF} \r^* \a_{\vec{k}} \right] \ket{0}.
\end{equation}
We note that the wavefunction \eqref{eq:MFansatz} is asymptotically exact in the limit of a localized impurity, i.e. when $J\rightarrow 0$. However, from the case of unconfined impurities it is known that the MF ansatz \eqref{eq:MFansatz} is unable to capture strong coupling physics \cite{Devreese2013} corresponding to the regime of very large interaction strength $g_\eff$ \cite{Cucchietti2006,Bruderer2007,Tempere2009,Casteels2011}.

To obtain self-consistency equations for the polaron ground state we minimize the MF  variational energy $\HMF$,
\begin{equation}
 \HMF=\bra{\Psi^\MF_q}\H_q \ket{\Psi^\MF_q} \stackrel{!}{=} \text{min}.
 \label{eq:HMFmin}
\end{equation}
As shown in Appendix \ref{sec:StatMFpolaron}, the MF energy functional can be written as
\begin{multline}
\HM[\alpha_{\vec{\kappa}}] = - 2 J e^{-C[\alpha_{\vec{\kappa}}]} \cos \l a q -S[\alpha_{\vec{\kappa}}] \r \\ +\int d^3k ~ \left[ \omega_k |\alpha_k|^2 + V_k \l \alpha_k + \alpha_k^* \r \right],
 \label{eq:HMfunctionalMainText}
\end{multline}
where we introduced the functionals
\begin{flalign}
 C[\alpha_{\vec{k}}] &= \int d^3 \vec{k} |\alpha_{\vec{k}}|^2 (1- \cos (a k_x)),  \label{eq:Cqdef} \\
 S[\alpha_{\vec{k}}] &= \int d^3 \vec{k} |\alpha_{\vec{k}}|^2 \sin (a k_x) . \label{eq:Sqdef}
\end{flalign}
Eq.\eqref{eq:HMFmin} together with \eqref{eq:HMfunctionalMainText} then yields the MF self-consistency equations for the polaron ground state,
\begin{equation}
\alpha_{\vec{k}}^\MF = -  V_k / \Omega_{\vec{k}}[\alpha_{\vec{\kappa}}^\MF],
\label{eq:MFselfCons}
\end{equation}
where we defined yet another functional
\begin{multline}
\Omega_{\vec{k}}[\alpha_{\vec{\kappa}}] = \omega_k + 2 J e^{- C[\alpha_{\vec{\kappa}}]}  \Bigl[ \cos \l a q  -S[\alpha_{\vec{\kappa}}] \r  \\ - \cos \l a q- a k_x -S[\alpha_{\vec{\kappa}}] \r \Bigr].
\label{eq:effPhononDispersion}
\end{multline}
This frequency $\Omega_{\vec{k}}[\alpha_{\vec{\kappa}}^\MF]$ can be interpreted as the renormalized phonon dispersion at total quasimomentum $q$. 

Importantly for numerical evaluation, Eq.\eqref{eq:MFselfCons} reduces to a set of only two self-consistency equations for $C^\MF=C[\alpha_{\vec{k}}^\MF]$ and $S^\MF=S[\alpha_{\vec{k}}^\MF]$: Plugging $\alpha_{\vec{k}}^\MF$ from \eqref{eq:MFselfCons} into the definitions \eqref{eq:Cqdef}, \eqref{eq:Sqdef} readily yields
\begin{flalign}
C^\MF &= \int d^3 \vec{k} ~ \left| \frac{V_k}{\Omega_{\vec{k}}(C^\MF,S^\MF)} \right|^2  \bigl( 1 - \cos (a k_x ) \bigr) \label{eq:selfConsC}, \\
S^\MF &= \int d^3 \vec{k} ~ \left| \frac{V_k}{\Omega_{\vec{k}}(C^\MF,S^\MF)} \right|^2  \sin (a  k_x ).  \label{eq:selfConsS}
\end{flalign}
Moreover, from the analytic form of $\Omega_{\vec{k}}$ Eq.\eqref{eq:effPhononDispersion} we find the following exact symmetries of the solution under spatial inversion $q \rightarrow -q$,
\begin{equation}
C^\MF(-q) = C^\MF(q), \qquad S^\MF(-q) = - S^\MF(q).
\end{equation}

\subsection{Results: equilibrium properties}
\label{subsec:polaronGSresults}

\begin{figure}[t]
\centering
\includegraphics[width=0.43\textwidth]{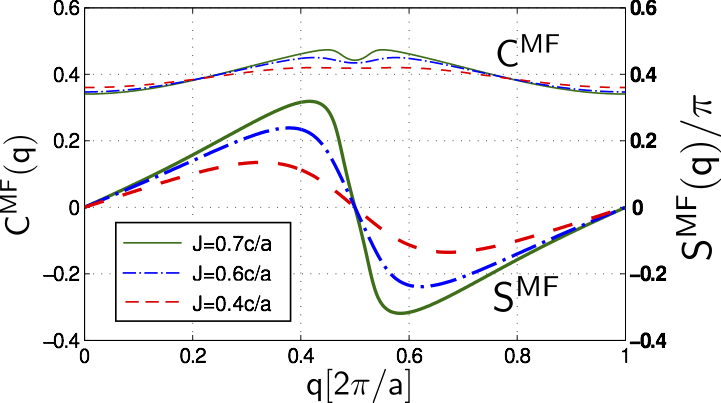}
\caption{(Color online) The MF  polaron ground state at total quasimomentum $q$ is characterized by $C^\MF(q)$ (upper thin lines) and $S^\MF(q)$ (lower thick lines). These quantities are plotted for various hoppings $J$, all in the subsonic regime. When approaching the transition towards supersonic polarons (which takes place slightly above $J=0.8 c/a$ in this case) the phase shift $S^\MF(q)$ develops a strong dispersion around $q=\pi/a$. At the same point a pronounced local minimum of $C^\MF(q)$ develops. We used $\xi=5 a, \lho=a/\sqrt{2}$ and $g_\eff=10$.}
\label{fig:staticMFpolaronCqSq}
\end{figure}

In FIG.\ref{fig:staticMFpolaronCqSq} we show the solutions $C^\MF$ and $S^\MF$ of the self-consistency equations \eqref{eq:selfConsC}, \eqref{eq:selfConsS} as a function of total quasimomentum $q$ for different hoppings. 
For weak interactions and not too close to the subsonic to supersonic transition  we find $S^\MF(q)\approx0$ while $C^\MF(q)\approx \text{const}$. In this limit the MF polaron dispersion becomes
\begin{equation}
\omega_\text{p}(q) \approx E_\text{b} - 2 J^*\cos(q a),
\label{eq:Jstar}
\end{equation}
cf.\eqref{eq:HMfunctionalMainText}. Here $J^* = J e^{-C^\MF}$ describes the renormalized hopping of the polaron, and we obtain a similar exponential suppression as reported in \cite{Bruderer2008}. $E_\text{b}$ describes the binding energy of the polaron.

\begin{figure}[t]
\centering
\includegraphics[width=0.5\textwidth]{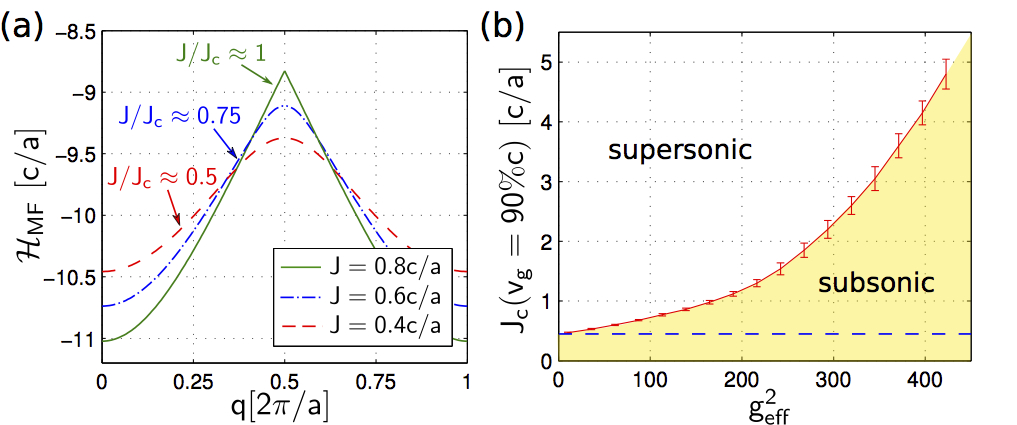}
\caption{(Color online) (a) MF  polaron dispersion $\HMF(q)$ for different impurity hoppings $J$, where the BEC MF shift $g_\IB n_0$ was neglected (it depends not only on the coupling strength $g_\eff$ but also on the BEC density $n_0$ which we did not specify here). For larger $J \gtrsim 0.8 c/a \approx J_\text{c}$ the group velocity $v_\text{g} = \partial_q \HMF(q)$ exceeds the speed of sound $c$ for some quasimomentum $q$. The interaction strength was $g_\eff=10$. (b) Critical hopping $J$ where the maximal group velocity $\max_q v_\text{g}(q)$ is $90 \%$ of $c$, as a function of interaction strength squared $g_\eff^2$. For large interactions $g_\eff \gg 1$ the hopping where the polaron becomes supersonic is much larger than in the non-interacting case (dashed line). Errorbars are due to the finite mesh-size used to raster parameter space. In both figures we have chosen $\xi=5 a$ and $\lho=a/\sqrt{2}$.}
\label{fig:staticMFpolaron}
\end{figure}

In FIG.\ref{fig:staticMFpolaron}(a) we show the full polaron dispersion relation. For substantial interactions $g_\eff=10$ chosen in FIG.\ref{fig:staticMFpolaron} we find a transition from a subsonic to a supersonic polaron around $J_c \approx 0.8c/a$. For hoppings close to this transition point the renormalized dispersion deviates markedly from the cosine shape familiar from bare impurities, and we observe strong renormalization at the edge of the BZ, $q=\pm \pi / a$. At the same time the overall energy is shifted substantially as a consequence of the dressing with high-energy phonons. 

%%%%%%%%%%%%%%%%%%%%%%%%%%%%%%%%%%%%%%%%%%%%%%%%%%%%%
%%%%%%%%%%%%%%%%%%%%%%%%%%%%%%%%%%%%%%%%%%%%%%%%%%%%%

In FIG.\ref{fig:staticMFpolaron}(b) we show the MF phase diagram. To this end we calculated the critical hopping $J_\text{c}$ where the maximal polaron group velocity in the BZ exceeds $90\%$ of the speed of sound $c$. (We only went to $90\%$ because close to the transition to supersonic polarons, solving the MF equations for $C[\alpha_{\vec{k}}^\MF]$ and $S[\alpha_{\vec{k}}^\MF]$ becomes increasingly hard numerically.) We observe that for large interactions the polaron is subsonic, even for bare hoppings $J$ one order of magnitude larger than the non-interacting critical hopping $J^{(0)}_\text{c}=c/2a$. This is in direct analogy to the strong mass renormalization predicted for free polarons, see e.g. \cite{Tempere2009,BeiBing2009,Shashi2014RF}. Interestingly we observe different behavior for weakly and strongly interacting polarons; We fitted the critical hopping to the curve
\begin{equation}
J_\text{c}(v_\text{g}=0.9c) = 0.9 J_\text{c}^{(0)} + g^2_\eff C_1 \l 1 + \l \frac{g_\eff}{g_\eff^c} \r^4 \r,
\end{equation}
varying parameters $C_1, g_\eff^c$. In this way we obtain a cross-over at $g_\eff^c=14.2$ for the parameters from FIG.\ref{fig:staticMFpolaron}.

%%%%%%%%%%%%%%%%%%%%%%%%%%%%%%%%%%%%%%%%%%%%%%%%%%%%%
%%%%%%%%%%%%%%%%%%%%%%%%%%%%%%%%%%%%%%%%%%%%%%%%%%%%%

%%%%%%%%%%%%%%%%%%%%%%%%%%%%%%%%%%%%%%%%%%%%%%%%%%%%%
\begin{figure}[b]
\centering
\includegraphics[width=0.34\textwidth]{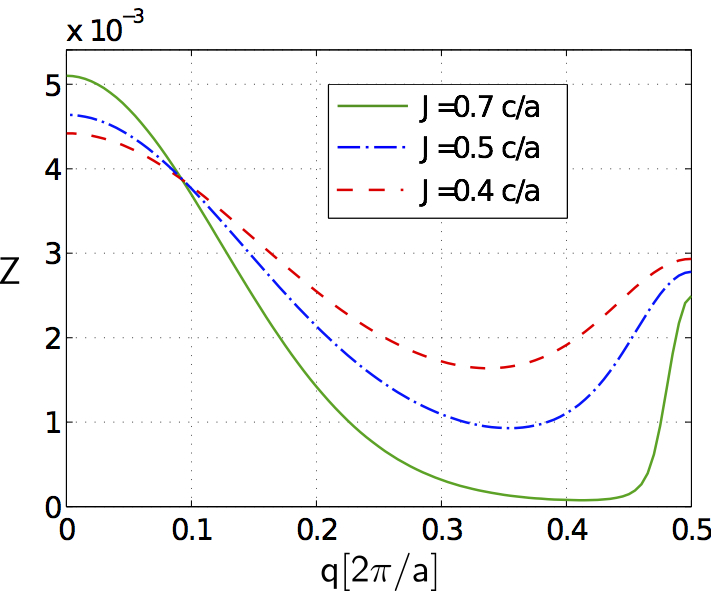}
\caption{(Color online) Dependence of the quasiparticle weight $Z$ of the polaron on quasimomentum $q$. We used the static MF polaron ground state to calculate $Z=Z_\MF$, which according to Eq.\eqref{eq:ZMFpolaron} is related to the average number of phonons in the polaron cloud, $Z_\MF=e^{-\langle N_{\text{ph}} \rangle }$. We have chosen $\xi=5 a, \lho=a/\sqrt{2}$ and $g_{\eff}=10$ as in FIGs.\ref{fig:staticMFpolaronCqSq} and \ref{fig:staticMFpolaron} (a).}
\label{fig:ZMFpolaron}
\end{figure}
%%%%%%%%%%%%%%%%%%%%%%%%%%%%%%%%%%%%%%%%%%%%%%%%%%%%%

We also consider the the quasiparticle weight $Z$, which is another quantity characterizing the polaron ground state. $Z$ is defined as the overlap between the bare and the dressed impurity state,
\begin{eqnarray}
 Z = |\langle 0 | \Psi_q \rangle|^2,
\end{eqnarray}
and can e.g. be measured using radio-frequency absorption spectoscopy of the impurity \cite{Shashi2014RF,Rath2013}. Within the MF approximation \eqref{eq:MFansatz} $\ket{\Psi_q} = \ket{\Psi_q^\MF}$, $Z$ is directly related to the number of phonons in the polaron cloud,
\begin{equation}
Z_\MF = \exp \l - \int d^3 \vec{k} ~ |\alpha^\MF_{\vec{k}}|^2\r = e^{- \langle N_{\text{ph}}\rangle}.
\label{eq:ZMFpolaron}
\end{equation}
Note, however, that this relation between the quasiparticle weight and the number of excited phonons is specific to the MF wavefunction and originates from its Poissonian phonon number statistics.

In FIG.\ref{fig:ZMFpolaron} the dependence of the MF quasiparticle weight on quasimomentum is shown. For the relatively strong coupling we have chosen, $Z \ll 1$ and the corresponding number of phonons is $N_{\text{ph}} = -\log(Z_\MF)$, taking values between $N_\ph = 5$ and $N_\ph = 9$ in the particular case of FIG.\ref{fig:ZMFpolaron}.
Importantly, we observe that the polaron properties are strongly quasimomentum dependent. Especially close to the subsonic to supersonic transition (i.e. for larger hopping $J$), we find an abrupt change of the quasiparticle weight close to the edge of the BZ. This is related to the peak observed in the renormalized polaron dispersion in FIG.\ref{fig:staticMFpolaron} (a). We interpret both these features as an onset of the subsonic to supersonic transition, which takes place at the edges of the BZ for strong impurity-boson interactions like in FIG.\ref{fig:staticMFpolaron}.

\section{Polaron Bloch oscillations and \\ adiabatic approximation}
\label{sec:adiabBO}
In this section we discuss how a uniform force acting on the impurity affects coherent polaron wavepacket dynamics. To this end we derive the equations of motion of a time-dependent variational state, and give an approximate solution using the adiabatic principle. From the latter we calculate real-space impurity trajectories. We close the section by pointing out how these trajectories can be used to measure the renormalized polaron dispersion in an experiment. In the following section we will check the validity of the adiabatic approximation by solving full non-equilibrium dynamics.

\subsection{Time-dependent variational wavefunctions}
We now treat the fully time-dependent Hamiltonian from Eq.\eqref{eq:polaronHam}, allowing us to solve for polaron dynamics. Our logic is as follows:  we decompose the wavefunction of the impurity-BEC system into different quasimomentum sectors, and use the conservation of quasimomentum of the polaron, which we established in Sec.~\ref{sec:LLP}, to treat each quasimomentum sector independently.

To this end, at time $t=0$, we consider a general initial wavefunction $\psi_j^\text{in}$ of the impurity\footnote{To be precise, $\psi_j^\text{in}$ denotes the projection of the initial impurity wavefunction $\psi_\text{I}^{\text{in}}(\vec{r})$ onto the $j^{\rm th}$ Wannier basis function $w_j(\vec{r})$, i.e.  $\psi^{\text{in}}_\text{I}(\vec{r}) = \sum_j \psi_j^{\text{in}} w_j(\vec{r})$} when the force is switched off, and for simplicity we assume complete absence of phonons. Thus, the initial quantum state reads
\begin{equation}
\ket{\Psi(0)} =  \sum_j \psi_j^\text{in} \cd_j \ket{0}_c \otimes \ket{0}_a,
\label{eq:PsiInitial}
\end{equation}
where $\ket{0}_c$ and $\ket{0}_a$ denote the impurity and phonon vacuum respectively. Note that Eq.\eqref{eq:PsiInitial} is true not only in the lab frame, but also in the polaron frame, i.e. after applying the Lee-Low-Pines transformation \eqref{eq:BOtransform}; Because in the absence of phonons we have $\ad_{\vec{k}} \a_{\vec{k}} \ket{\Psi(0)} =0$, for the initial state from Eq.\eqref{eq:PsiInitial} it holds $\hat{S} \ket{\Psi(0)}=\ket{\Psi(0)}$.

The initial state \eqref{eq:PsiInitial} considered in most of the remaining part of this paper can be realized experimentally by different means. For instance, if Feshbach resonances are used to realize strong impurity-boson interactions one can quickly change the magnetic field strength from a value far away from the resonance to a value very close to it at time $t=0$. Therefore an initially non-interacting impurity, immersed in a cold BEC, suddenly starts to interact strongly with the surrounding phonons as the magnetic field approaches the Feshbach resonance. 

Alternatively, if a different internal (e.g. hyperfine) state of the majority bosons is used as an impurity like e.g. in \cite{Fukuhara2013}, the initial state can be prepared by applying a microwave pulse, which is possible also in combination with local addressing techniques \cite{Weitenberg2011,Fukuhara2013}. In this case, however, the preparation of a phonon vacuum state like in Eq.\eqref{eq:PsiInitial} is hard to achieve since a spin-flip always comes along with a local excitation of the BEC. Nevertheless, the true initial state for this situation can be calculated exactly if after a local spin-flip the impurity is tightly confined by an addressing beam \cite{Weitenberg2011,Fukuhara2013} until the dynamic evolution is started at time $t=0$. In fact, a sufficiently tight local confinement of the impurity corresponds to vanishing hopping $J=0$, and in this case the MF ansatz Eq.\eqref{eq:MFansatz} yields the exact phonon ground state with coherent state amplitudes $\alpha_{\vec{k}}^{(J=0)}$. Therefore, assuming the system has enough time to relax to its ground state after preparation of the tightly confined impurity on the central site $j=0$, the initial state reads
\begin{equation}
\ket{\Psi(0)} =  \cd_0 \ket{0}_c \otimes \prod_{\vec{k}} \ket{\alpha_{\vec{k}}^{(J=0)}}.
\label{eq:PsiInitial2}
\end{equation}
Like the state from Eq.\eqref{eq:PsiInitial} this wavefunction is invariant under the Lee-Low-Pines transformation \eqref{eq:BOtransform}, but in this case because of a trivial action of the impurity position operator, $\hat{X} \ket{\Psi(0)}=0$.

Next, focusing on Eq.\eqref{eq:PsiInitial} again for concreteness, we decompose the initial state into its different quasimomentum sectors, which is achieved by taking a Fourier-transform of the impurity wavefunction,
\begin{equation}
 f_q = \frac{1}{\sqrt{L/a}} \sum_j e^{i q a j} \psi_j^\text{in}.
 \label{eq:fqDefFT}
\end{equation}
When the force is switched on at time $t=0$, all quasimomentum sectors evolve individually without any couplings between them. As a consequence the amplitudes $f_q$ defined above are conserved, and we may write the time-evolved quantum state in the polaron frame as
\begin{equation}
\ket{\Psi(t)} =  \sum_{q \in \BZ} f_q \cd_q \ket{0}_c \otimes \ket{\Psi_q(t)}.
\end{equation}

At given initial quasimomentum $q(0)=q$ and for finite driving force $F$ we can make a variational ansatz for the phonon wavefunction similar to the MF case Eq.\eqref{eq:MFansatz}, but with time-dependent parameters,
\begin{equation}
\ket{\Psi_q(t)} = e^{-i \chi_q(t)} \prod_{\vec{k}} \ket{\alpha_{\vec{k}}(t)}.
\label{eq:DMFvarState}
\end{equation}
To derive equations of motion for $\alpha_{\vec{k}}(t)$ we use Dirac's time dependent variational principle and arrive at (for details see Appendix \ref{sec:DiracVarPrinc})
\begin{equation}
 i \partial_t \alpha_{\vec{k}}(t) =\Omega_{\vec{k}}[\alpha_{\vec{\kappa}}(t)] ~  \alpha_{\vec{k}}(t) + V_k.
\label{eq:DMF_EOM}
\end{equation}
Here $\Omega_{\vec{k}}[\alpha_{\vec{\kappa}}(t)]$ is the renormalized phonon dispersion, see Eq.\eqref{eq:effPhononDispersion}, but evaluated for time-dependent $\alpha_{\vec{\kappa}}(t)$. Note that $\Omega_{\vec{k}}$ explicitly depends on $q(t)=q - Ft$. In Appendix \ref{sec:DiracVarPrinc} we also derive an equation describing the dynamics of the global phases $\chi_q(t)$,
\begin{equation}
\partial_t \chi_q = \frac{i}{2} \int d^3\vec{k} \l \dot{\alpha}_{\vec{k}}^*\alpha_{\vec{k}} - \dot{\alpha}_{\vec{k}} \alpha_{\vec{k}}^* \r+ \prod_{\vec{k}} \bra{\alpha_{\vec{k}}} \H_{q(t)}\ket{\alpha_{\vec{k}}}.
\label{eq:DMF_EOM_phases}
\end{equation}

\subsection{Adiabatic approximation}
Before presenting the full numerical solutions of Eqs.\eqref{eq:DMF_EOM}, \eqref{eq:DMF_EOM_phases}, we first discuss the adiabatic approximation. It assumes that the polaron follows its ground state without creating additional excitations, i.e. without emission of phonons. We may thus approximate the dynamical phonon wavefunction by
 \begin{equation}
\ket{\Psi_q(t)} \approx e^{-i \chi_q(t)}\ket{\Psi^\MF_{q(t)}}.
\label{eq:adiabApprx}
\end{equation}
The intuition here is that the time-scale for polaron formation is much faster than the dynamics of BO. In particular, $\ket{\Psi^\MF_{q(t)}}$ is simply the equilibrium polaron MF solution for quasimomentum $q(t)$ obtained in Sec.~\ref{sec:StaticProps}, which changes in time according to
\begin{equation}
q(t) = q(0) - Ft.
\label{eq:qtq0Ft}
\end{equation}
Additionaly, we allow for a time-dependence of the global phase, which we obtain from Eq.\eqref{eq:DMF_EOM_phases},
\begin{equation}
\chi_q(t) = \int_0^t dt' ~ \HMF(q(t')).
\end{equation}

\subsection{Polaron trajectory}
\label{subsec:PolaronTrajectory}
Next, we derive the real-space trajectory of the polaron. To this end we calculate the impurity density, which can be expressed as
\begin{equation}
\left\langle \cd_j \c_j \right\rangle = \frac{1}{L/a} \sum_{q_2,q_1 \in \BZ} e^{i a (q_2 - q_1) j} A_{q_2,q_1}(t) f^*_{q_2} f_{q_1}.
 \label{eq:njtDMF}
\end{equation}
This formula is derived in Appendix \ref{sec:ImpDensity}, and it requires knowledge of the time-dependent overlaps
\begin{equation}
A_{q_2,q_1}(t)=\bra{\Psi_{q_2}(t)} \Psi_{q_1}(t) \rangle.
\label{eq:defNonEqGreen}
\end{equation}
They consist of two factors, $A_{q_2,q_1}=\mathcal{A}_{q_2,q_1} \mathcal{D}_{q_2,q_1}$; The phases obey $|\mathcal{A}_{q_2,q_1}|=1$ and are given by
\begin{equation}
\mathcal{A}_{q_2,q_1}(t) = \exp \left[ i \l \chi_{q_2}(t) - \chi_{q_1}(t) \r \right],
\label{eq:cohPart}
\end{equation}
whereas the amplitudes $\mathcal{D}_{q_2,q_1}$, determined by phonon dressing, are
\begin{equation}
\mathcal{D}_{q_2,q_1} = \prod_{\vec{k}} \bra{\alpha_{\vec{k}}(q_2,t)} \left. \alpha_{\vec{k}}(q_1,t) \right\rangle.
\label{eq:incohPart}
\end{equation}
Within the adiabatic approximation we set $\alpha_{\vec{k}}(q,t) = \alpha^\MF_{\vec{k}}(q(t))$. For non-interacting impurities the phases alone give rise to BO, while the amplitude is trivial $\mathcal{D}=1$. When interactions of the impurity with the phonon bath are included, $| \mathcal{D}| <1$ and interference is suppressed.

To get an insight into the BO of polarons we begin by discussing a special case of a polaron wavepacket prepared with narrow distribution in quasimomentum space. In particular we will consider an initial ground state polaron wavepacket centered around $q=0$, which is described by
\begin{equation}
\ket{\Psi(0)} = \sqrt{ \frac{2 L_\I}{\sqrt{2 \pi}} } \sum_{q\in \BZ}   e^{-q^2 L_\I^2} \cd_q \ket{0}_c \otimes \ket{\Psi^\MF_q},
\label{eq:PsiIniAdiabAprx}
\end{equation}
and where $L_\I$ denotes its width in real space. We will assume $L_\I \gg a$ in the analysis below, such that all wavepackets carry a well-defined quasimomentum. Therefore, in Eq. \eqref{eq:njtDMF} only neighboring momenta $|q_2-q_1|\ll2 \pi/a$ contribute, allowing us to expand the exponent of $A_{q_2,q_1}$ to second order in $|q_2-q_1|$. In this way we obtain the adiabatic impurity density (the detailed calculation can be found in Appendix \ref{sec:AdiabaticWavePacket})
\begin{equation}
n(x,t) = e^{ - \frac{(x-X(t))^2}{2 \l L_\I^2 + \Gamma^2(t) \r } } \left[ 2 \pi \l L_\I^2 + \Gamma^2(t) \r \right]^{-1/2}.
\label{eq:adiabImpDsty}
\end{equation}
Note that due to the large spatial extent assumed for the polaron wavepacket we treated $a j = x$ as a continuous variable here. 

The center-of-mass coordinate of the polaron is determined by $\mathcal{A}_{q_2,q_1}$ and it reads
\begin{equation}
X(t) = X(0) + \left[ \HMF(F t) - \HMF(0) \right] / F.
\label{eq:XtAd}
\end{equation}
The amplitude $\mathcal{D}_{q_2,q_1}$, meanwhile, leads to \emph{reversible} broadening of the polaron wavepacket,
\begin{equation}
\Gamma^2(t) = \int d^3 \vec{k} \l \left. \partial_q \alpha_k^\MF \right\vert_{q=-F t} \r^2.
\label{eq:GtAd}
\end{equation}
From Eq.\eqref{eq:XtAd} we thus conclude that a measurement of the polaron center $X(t)$ directly reveals the renormalized polaron dispersion relation. Although derived from a simplified theory, we expect that this result holds more generally beyond MF approximation of the polaron ground state.

\section{Non-adiabatic corrections}
\label{sec:Dynamics}
In this section we study the full non-equilibrium dynamics of the driven polaron by numerically solving for the time-dependent MF wavefunction Eq.\eqref{eq:DMFvarState}. We start from the phonon vacuum and some initial impurity wavefunction $\psi_j^{\text{in}}$, see Eq.\eqref{eq:PsiInitial}, mostly chosen to be a Gaussian wavepacket with a width $L_\I$ of several lattice sites and vanishing mean quasimomentum $q=0$. After switching on the impurity-boson interactions at time $t=0$, we find polaron formation and discuss the validity of the adiabatic approximation for a description of the subsequent dynamics (in \ref{subsec:BOcorrections}). We also briefly discuss the case of initially localized impurities (in \ref{sec:BeyondWavePacket}).

To solve equations of motion \eqref{eq:DMF_EOM} we employ spherical coordinates $k,\vartheta,\phi$ and make use of azimuthal symmetry around the direction of the impurity lattice. We introduce a grid in $k-\vartheta$ space (typically $170 \times 40$ grid points) and use a standard matlab solver for ordinary differential equations. From the so-obtained solutions $\alpha_{k,\vartheta}(t)$ and $\chi_q(t)$ we calculate $A_{q_2,q_1}(t)$ using Eqs.\eqref{eq:cohPart}, \eqref{eq:incohPart}, giving access to impurity densities for arbitrary impurity initial conditions, see Eq.\eqref{eq:njtDMF}.

 \subsection{Impurity dynamics beyond the adiabatic approximation}
 \label{subsec:BOcorrections}
 
 %%%%%%%%%%%%%%%%%%%%%%%%%%%%%%%%%%%%%%%%%%%%%%%%%%%%%
\begin{figure}[b]
\centering
\includegraphics[width=0.42\textwidth]{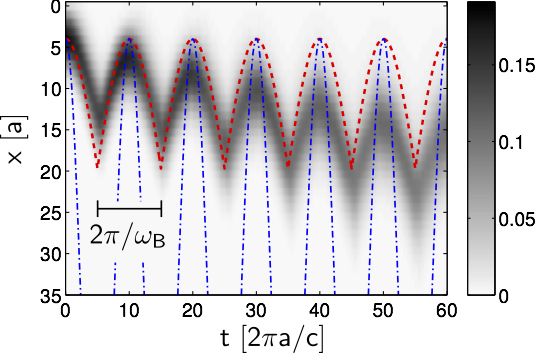}
\caption{(Color online) Impurity density $\langle \cd_j \c_j \rangle$ (color code) with $j a = x$ for a heavily dressed impurity. The polaron dynamics, starting from phonon vacuum, is compared to the result from the adiabatic approximation (red, dashed) as well as the trajectory of a non-interacting impurity wavepacket (blue, dashed-dotted). The parameters are $J=1.7 c/a$, $F=0.1 c/a^2$, $g_{\eff}=17.32$, $\lho=a/\sqrt{2}$ and $\xi=5a$.}
\label{fig:SM_adiabApprxOld1}
\end{figure}
%%%%%%%%%%%%%%%%%%%%%%%%%%%%%%%%%%%%%%%%%%%%%%%%%%%%%
 
To extend our analysis beyond the assumption that the system follows its ground state adiabatically, we now consider the full dynamical equations \eqref{eq:DMF_EOM} and \eqref{eq:DMF_EOM_phases}. We assume that the system starts in the initial state \eqref{eq:PsiInitial} with the phonons in their vacuum state, and at time $t=0$ interactions between the impurity and the bosons are switched on abruptly. We chose the initial impurity wavefunction $\psi_j^{\text{in}}$ to be a Gaussian wavepacket (standard deviation $L_\text{I}$) like in the discussion of the adiabatic approximation, see Sec. \ref{subsec:PolaronTrajectory}. Thus the amplitudes $f_q$ read $f_q = e^{-(q L_\text{I})^2} (2 L_\text{I})^{1/2} (2 \pi)^{-1/4}$, as in Eq.\eqref{eq:fqDefFT}. The global phases vanish initially, i.e. we set $\chi_q(0)=0$ for all quasimomenta $q$.

In FIG.\ref{fig:SM_adiabApprxOld1} the evolution of the impurity density is shown for a strongly interacting case. Although the impurity hopping $J=1.7 c/a$ exceeds the critical hopping $J_c^{(0)}=0.5 c/a$ where a bare particle becomes supersonic by more than a factor of three, we observe well defined BO with group velocities of the wavepacket below the speed of sound $c$. By investigating the mean phonon number we moreover find that polaron formation takes place on a time-scale $\xi/c$ after which a quasi steady state is reached. 

 %%%%%%%%%%%%%%%%%%%%%%%%%%%%%%%%%%%%%%%%%%%%%%%%%%%%%
\begin{figure}[t]
\centering
\includegraphics[width=0.435\textwidth]{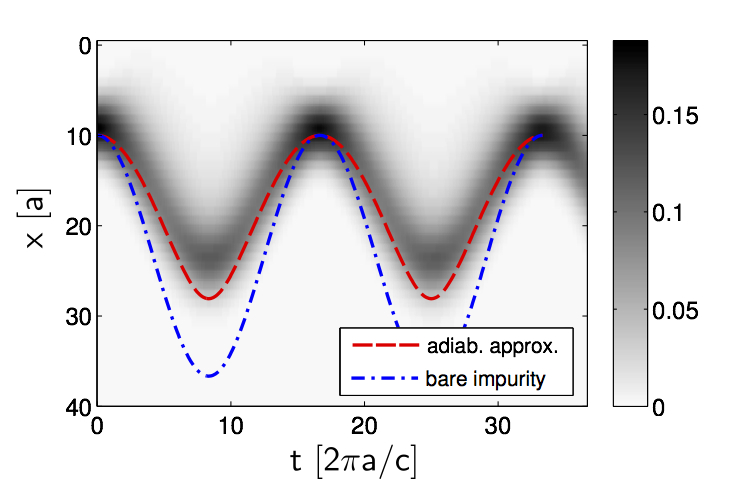}
\caption{(Color online) Impurity density $\langle \cd_j \c_j \rangle$ (color code) with $j a = x$ for a weakly driven polaron. For comparison the trajectory of a non-interacting impurity wavepacket is shown (dashed-dotted). The polaron dynamics is well described by the adiabatic approximation (dashed), which in turn is given by the polaron dispersion relation, see Eq.\eqref{eq:XtAd}. Thus direct imaging of the impurity density allows a measurement of the polaron dispersion. The parameters are $J=0.4 c/a$, $F=0.06 c/a^2$, $g_{\eff}=10$, $\lho=a/\sqrt{2}$ and $\xi=5a$.}
\label{fig:SM_adiabApprx}
\end{figure}
%%%%%%%%%%%%%%%%%%%%%%%%%%%%%%%%%%%%%%%%%%%%%%%%%%%%%

Along with the plot in FIG.\ref{fig:SM_adiabApprxOld1} we show the result of the adiabatic approximation. Although the latter can not capture the initial polaron formation, it is expected to be applicable once a steady state is reached \footnote{After the quench there is excess energy which will however be carried away by phonons. When tracing out these emitted phonons, we expect the remaining state to be well described by a ground state polaron, provided that equilibration mechanisms are available.}. In the case shown in the figure, however, non-adiabatic corrections play an important role and we observe a pronounced polaron drift in the direction of the force $F$. Moreover irreversible broadening of the polaron wavepacket takes place. Nevertheless the shape of the BO trajectory, including its pronounced peaks and the amplitude of oscillations, can be understood from the adiabatic result. For smaller hopping and smaller interactions the adiabatic approximation compares even better with the full numerics, as is shown in FIG.\ref{fig:SM_adiabApprx}.

%%%%%%%%%%%%%%%%%%%%%%%%%%%%%%%%%%%%%%%%%%%%%%%%%%%%%
\begin{figure}[t]
\centering
\includegraphics[width=0.48\textwidth]{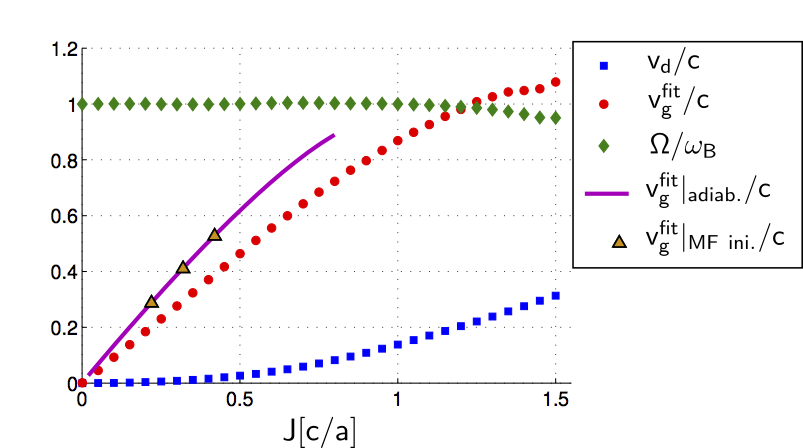}
\caption{(Color online) The impurity center-of-mass $\overline{X}(t)$ obtained from our time-dependent variational calculation can be fitted to the expression from  Eq.\eqref{eq:FitPolTraj}. The dependencies of the fitting parameters $v_{\text{d}}$, $v_\text{g}^{\text{fit}}$ as well as $\Omega$ on the hopping strength $J$ are shown in this figure. In the subsonic regime ($J \lesssim 1.2 c/a$) the fitted maximum group velocity $v_\text{g}^\text{fit}$ (red bullets) is compared to the result obtained from the adiabatic approximation (solid line). To this end we fitted the polaron trajectory obtained from the adiabatic approximation to the same curve from Eq. \eqref{eq:FitPolTraj} and plotted the so-obtained velocity $v_{g}^{\text{fit}} |_\text{adiab.}$. The observed deviations of our data from the adiabatic theory can be explained by the initial quench: when starting the dynamics from the MF polaron ground state (instead of a non-interacting impurity) the resulting trajectory $v_\text{g}^{\text{fit}} |_{\text{MF ~ini.}}$ is in excellent agreement with our theoretical prediction (triangles $\triangle$). The parameters were $F=0.2 c/a^2, g_\text{eff}=10, \xi = 5 a$ and $\lho = a/\sqrt{2}.$}
\label{fig:adiabApprxVarJ}
\end{figure}
%%%%%%%%%%%%%%%%%%%%%%%%%%%%%%%%%%%%%%%%%%%%%%%%%%%%%

To perform a more quantitative analysis when adiabaticity may be assumed, we determine the center-of-mass $\overline{X}(t) = \sum_j  j \langle  \cd_j \c_j \rangle$ of the impurity wavefunction from the full variational calculation and fit it to
\begin{equation}
 \overline{X}(t) = \frac{v_\text{g}^\text{fit}}{\Omega}  \cos \l \Omega t + \varphi \r  + v_\text{d} t + X_0.
\label{eq:FitPolTraj}
\end{equation}
Here $v_\text{g}^\text{fit}$ denotes the maximum polaron velocity in the absence of a drift. In FIG.\ref{fig:adiabApprxVarJ} the resulting fit parameters are shown as a function of the bare hopping $J$. We compare the value of $v_\text{g}^\text{fit}$ to the polaron group velocity expected from adiabatic approximation $v_\text{g}^\text{fit} |_\text{adiab.}$. The latter is obtained by fitting Eq.\eqref{eq:FitPolTraj} to the adiabatic trajectory. While the adiabatic theory captures correctly the qualitative behavior, on a quantitative level it overestimates the group velocity. This, however, is related to our initial conditions and not to a shortcoming of the adiabatic approximation in general. When starting the dynamics from the MF polaron state Eq.\eqref{eq:PsiIniAdiabAprx} instead of considering an interaction quench of the impurity, we find excellent agreement, with deviations below $1 \%$. This is demonstrated by a few data points in FIG.\ref{fig:adiabApprxVarJ}. The quench, on the other hand, leads to the creation of phonons, which are also expected to contribute to the dressing of the impurity in general \cite{Bruderer2007}.

Close to the subsonic to supersonic transition around $J_c\approx c/a$, the polaron drift velocity takes substantial values of $\approx 0.2 c$. We also note that, in the entire subsonic regime, the fitted BO frequency $\Omega$ is precisely given by the bare-impurity value $\omega_\text{B}$ (to within $<0.5 \%$ in the numerics). However, once the polaron becomes supersonic we observe a decrease of the frequency to $\Omega < \omega_\text{B}$. We attribute this effect to the spontaneous emission of phonons in regions of the BZ where the polaron becomes supersonic. Along with phonon emission comes emission of net phonon momentum $\Delta q_\text{ph}$, which has to be replenished by the external driving force, $\Delta q_\text{ph} = F\Delta t$. Thus an extra time $\Delta t$ is required for each Bloch cycle and as a consequence we expect the BO frequency of the polaron to decrease.

Within the adiabatic approximation we have shown that the wavepacket trajectory $X(t)$ allows a direct measurement of the renormalized polaron dispersion. We found that even when non-adiabatic effects are appreciable, the polaron dispersion can be reconstructed. BO can therefore be used as a \emph{tool} to measure polaronic properties, which are of special interest in the strongly interacting regime. We emphasize that our scheme does not rely on the specific variational method used above. As long as the ground state of the impurity interacting with the phonons of the surrounding BEC, is described by a stable polaron band, the real-space BO trajectory maps out the integrated group velocity, i.e. the band structure itself.

\subsection{Beyond wavepacket dynamics}
\label{sec:BeyondWavePacket}
%%%%%%%%%%%%%%%%%%%%%%%%%%%%%%%%%%%%%%%%%%%%%%%%%%%%%
\begin{figure}[t]
\centering
\includegraphics[width=0.5\textwidth]{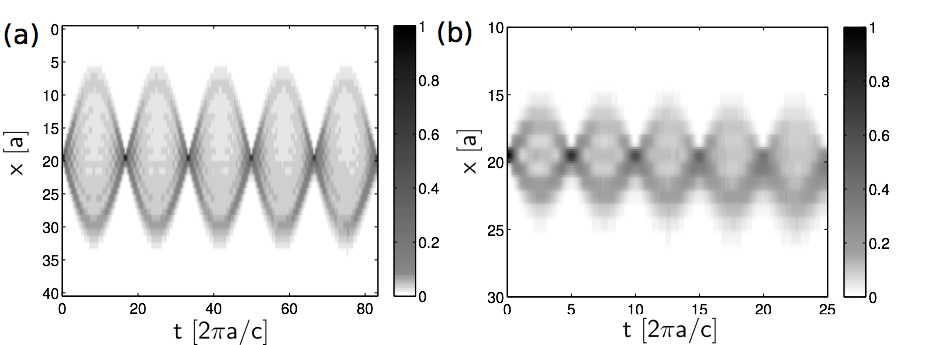}
\caption{(Color online) Impurity density $\langle \cd_j \c_j \rangle$ (color code) with $j a = x$ for an initially localized state on a single lattice site ($x(0)=20 a$ in this concrete example). (a) Weak driving $F=0.06 c/a^2$ and (b) stronger driving $F=0.2 c/a^2$. Other parameters are $J=0.3 c/a$, $g_\eff=10$, $\xi=5 a$ and $\lho=a/\sqrt{2}$ in both cases.}
\label{fig:SM_beyWavePack}
\end{figure}
%%%%%%%%%%%%%%%%%%%%%%%%%%%%%%%%%%%%%%%%%%%%%%%%%%%%%
Motivated by their possible application for measurements of the renormalized dispersion, we focused on polaron wavepackets so far. Our variational treatment, however, is applicable to any initial wavefunction. In FIG.\ref{fig:SM_beyWavePack} we show two examples starting from an impurity which is localized on a single lattice site, still assuming phonon vacuum initially. Since all momenta are occupied, we first observe interference patterns which are symmetric under spatial inversion $x \rightarrow -x$. For large enough interactions and sufficiently strong driving however, we observe diffusion of the polaron and the interference patterns disappear. The maximum impurity density drops substantially and the symmetry under spatial inversion is lost. Moreover we observe a finite drift velocity of the polaron.

\section{Polaron transport}
\label{sec:NonAdiabatic}
In this section we discuss the polaron drift velocity $v_\text{d}$, which is the most important non-adiabatic effect and can also be interpreted as a manifestation of incoherent transport. After some brief general remarks about the problem, we present our numerical results for the current-force relation $v_\text{d}(F)$. These are obtained, like in the last section, from the time-dependent variational MF ansatz Eq.\eqref{eq:DMFvarState}, requiring numerical solutions of Eqs.\eqref{eq:DMF_EOM}, \eqref{eq:DMF_EOM_phases}. Next we derive a closed, semi-analytical expression for the current-force relation $v_\text{d}(F)$ from first principles in the limit of small polaron hopping $J^*=J e^{-C^\MF}$ and show that our predictions are in good quantitative agreement with the full time-dependent MF numerics. As a result, we find that the polaron drift in the weak-driving limit strongly depends on the dimensionality of the system. At the end of this section we discuss the connection between our results and the Esaki-Tsu relation, which originates from a purely phenomenological model of incoherent transport in a lattice potential. We find that, in the polaron case, this simplified model is unable to capture many key features of our findings. In particular it completely fails in the weak-driving regime and predicts a wrong dependence on the hopping strength $J$. 

%%%%%%%%%%%%%%%%%%%%%%%%%%%%%%%%%%%%%%%%%%%%%%%%%%%%%
\begin{figure*}[t]
\centering
\includegraphics[width=0.95\textwidth]{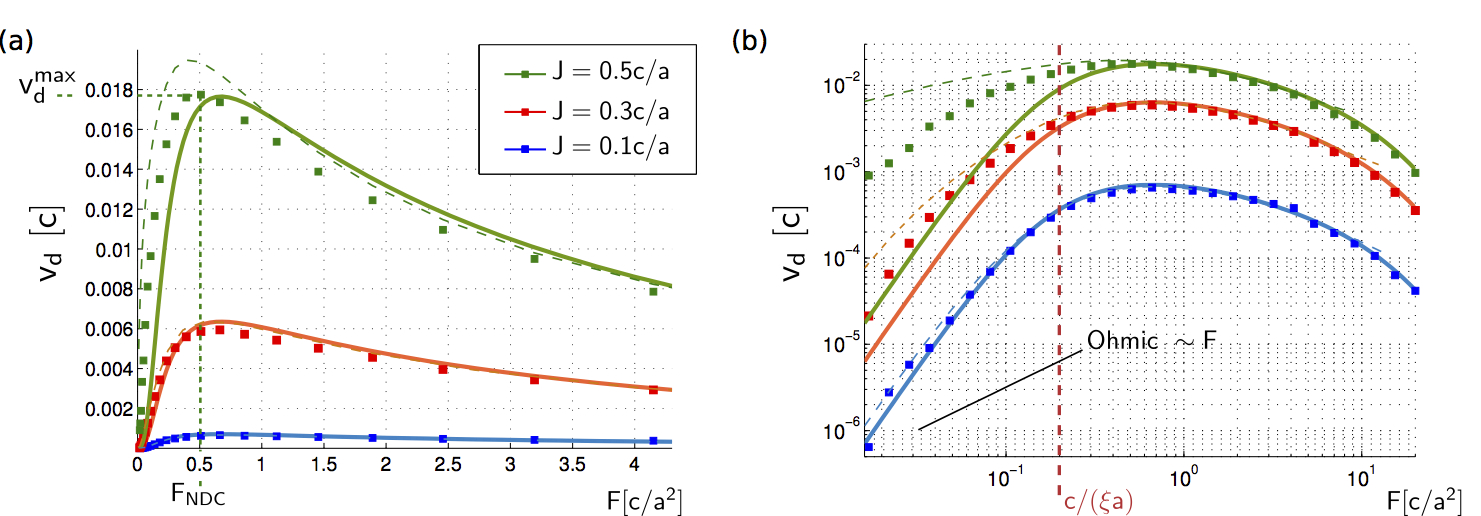}
\caption{(Color online) (a) Dependence of the polaron drift velocity $v_\text{d}$ (obtained from the fit Eq.\eqref{eq:FitPolTraj}) on the driving force $F$, for interaction strength $g_{\text{eff}}=3.16$ and various hoppings (top: $J=0.5c/a$, middle: $J=0.3c/a$, bottom: $J=0.1 c/a$). All curves show the same qualitative features: for small force $F$ the polaron current increases with $F$, it reaches its maximum $v_\text{d}^{\text{max}}$ at the negative differential conductance peak at $F_{\NDC}$ and for stronger driving $F > F_\NDC$ negative differential conductance $d v_\text{d} / d F < 0$ is observed. For each $J$ we also show the result from our analytical model Eq.\eqref{eq:gmaPh} of polaron transport (solid lines), free of any fitting parameters. We find excellent agreement for $J=0.3 c/a$ and $J=0.1 c/a$. We also plotted the prediction of an extended model (dashed lines, solution of the truncated Hamiltonian Eq.\eqref{eq:HflucMF}), which for $J=0.5 c/a$ yields somewhat better results.
 In (b) we show the same data (legend from (a) applies), but in double-logarithmic scale. In the lower left corner we indicated an Ohmic power-law dependence $\sim F$ (thin solid line). Comparison to our data shows a sub-Ohmic current-force dependence in the weak driving regime (the approximate power-laws have exponents in a range between $1$ and $3$). It starts roughly when $\omega_\text{B} = a F < c / \xi$, which is indicated by a dashed vertical line in (b).
For all curves we used $\xi=5a$ and $\lho = a/\sqrt{2}$ and simulated at least three periods of BO assuming an initial Gaussian impurity wavepacket with a width $L_\text{I}$ of three lattice sites.}
\label{fig:vd_F}
\end{figure*}
%%%%%%%%%%%%%%%%%%%%%%%%%%%%%%%%%%%%%%%%%%%%%%%%%%%%%

\subsection{General observations}
The fundamental Hamiltonian \eqref{eq:Hfund} is manifestly time-independent, and thus total energy is conserved. When the impurity slides down the optical lattice, the loss of potential energy $\dot{E}_{\text{pot}}=-F v_\text{d}$ requires a gain of radiative energy $E_\gamma$ in the form of phonons, $\dot{E}_\gamma = F v_\text{d}$. (This relation can also formally be derived from Eq.\eqref{eq:Hefft}.) Therefore the non-zero drift velocity of the polaron wavepacket observed e.g. in FIG.\ref{fig:setup}(b) comes along with phonon emission, albeit its velocity never exceeds the speed of sound $c$. Such phonon emission is not in contradiction to Landau's criterion for superfluidity, which is appropriate only for impurities (or obstacles in general) in a superfluid moving with a \emph{constant} velocity. However, the system considered here is driven by an external force $F$ which gives rise to periodic oscillations of the net quasimomentum of the system $q(t)$. We thus expect phonons to be emitted at multiples of the BO frequency $\omega=n \omega_{\text{B}}$, with rates $\gamma_\text{ph}(n \omega_{\text{B}})$. Using $\dot{E}_\gamma = \sum_n n \omega_\text{B} \gamma_\text{ph}(n \omega_{\text{B}})$, we can express the drift velocity as
\begin{equation}
v_\text{d} = a \sum_n n \gamma_\text{ph}(n \omega_{\text{B}}).
\label{eq:vdFstHarm}
\end{equation}

\subsection{Numerical results}
\label{subsec:numRes}
In FIG.\ref{fig:vd_F} we present numerical results for the current-force dependence at different hopping strengths $J$, in linear (a) and double-logarithmic scale (b). These curves were obtained by solving for the variational time-dependent MF wavefunction \eqref{eq:DMFvarState}. Like in the last section, we started from phonon vacuum and assumed a zero-quasimomentum impurity wavepacket extending over a few lattice sites. The center-of-mass $\overline{X}(t)$ of the resulting polaron trajectory was then fitted to Eq.\eqref{eq:FitPolTraj} from which $v_\text{d}$ was obtained as a fitting parameter.

All curves have a similar qualitative form: For small force $\omega_\text{B} \lesssim c/\xi$ the polaron current increases monotonically with $F$. Somewhere around $\omega_{\text{B}} \approx c/\xi$ the curvature changes and the polaron drift velocity takes its maximum value $v_\text{d}^{\text{max}}$ for a force $F_{\NDC}$. For even larger driving $\omega_{\text{B}}$ we find negative differential conductance, defined by the condition $d v_\text{d}/ dF < 0$. The maximum is also referred to as negative differential conductance peak. Previously all these features have been predicted by different polaron models for impurities in 1D condensates \cite{Bruderer2008,Bruderer2010}.

From the double-logarithmic plot in FIG.\ref{fig:vd_F} (b) we observe a \emph{sub-Ohmic} behavior in the weak driving regime. For the smallest achievable forces $F$, we can approximate our curves by power-laws $v_\text{d} \sim F^\gamma$. The observed exponents in FIG.\ref{fig:vd_F} (b) are in a range $\gamma=3.0$ (for $J=0.3 c/a$, $g_{\eff}=3.16$) to $\gamma=1.5$ (for $J=0.5 c/a$, $g_{\eff}=3.16$). While this behavior is clearly sub-Ohmic, it is hard to estimate how well these power-laws extrapolate to the limit $F \rightarrow 0$. Going to even smaller driving is costly numerically, because the required total simulation time for a few Bloch cycles $T \sim 1/F$ becomes large.

To our knowledge, the sub-Ohmic behavior in the weak-driving regime was not previously observed. As we discuss at the end of this section, it goes beyond the phenomenological Esaki-Tsu model for incoherent transport in lattice models. We show in the following that it is moreover tightly linked to the dimensionality $d$ of the condensate providing phonon excitations. For 1D systems, which were studied in some depth in the literature \cite{Bruderer2008,Bruderer2010,Johnson2011}, we do in fact expect Ohmic behavior for $F \rightarrow 0$. This is in agreement with the results of \cite{Bruderer2008,Bruderer2010,Johnson2011}.

\subsection{Semi-analytical current-force relation}
\label{subsubsec:analytics}
Now we want to extend our formalism used to describe the static polaron ground state in Sec.\ref{sec:StaticProps} by including quantum fluctuations. To this end we apply the following unitary transformation
\begin{equation}
\hat{U}(q) = \prod_{\vec{k}} \exp \l \alpha_{\vec{k}}^\MF(q) \ad_{\vec{k}} - \l \alpha_{\vec{k}}^\MF (q)\r^* \a_{\vec{k}} \r,
\label{eq:UMFfluc}
\end{equation}
where in the new frame $\a_{\vec{k}}$ describes quantum fluctuations around the MF solution in the absence of driving, $F=0$. In the case of a non-vanishing force $F \neq 0$, we can analogously obtain corrections to the adiabatic MF polaron solution Eq.\eqref{eq:DMFvarState}. To this end we have to make the transformation \eqref{eq:UMFfluc} time-dependent, $\hat{U}(t) := \hat{U}(q(t))$, where $q(t) = q(0) - F t$ (see Eq.\eqref{eq:qtq0Ft}).

By applying $\hat{U}(q(t))$, defined by Eq.\eqref{eq:UMFfluc} above, to the polaron Hamiltonian Eq.\eqref{eq:Hefft} we obtain the following time-dependent Hamiltonian describing quantum fluctuations around the adiabatic MF polaron solution in the case of a $d$-dimensional condensate,
\begin{multline}
\tilde{\mathcal{H}}(t) = \int d^d \vec{k} ~ \Omega_{\vec{k}}(q(t)) \ad_{\vec{k}} \a_{\vec{k}} + \mathcal{O}(J^* \a^2)
\\+ i F \int d^d \vec{k} ~ \l \partial_q \alpha^\MF_{\vec{k}} (q(t)) \r \left[  \ad_{\vec{k}} -   \a_{\vec{k}}  \right].
\label{eq:HflucMF}
\end{multline}
Here we introduced $J^*(q(t)):=J \exp \l - C^\MF(q(t)) \r$ and $\mathcal{O}(J^* \a^2)$ denotes terms describing corrections to the adiabatic solution beyond the MF description of the polaron ground state. The leading order terms have a form $ \sim J^* \a_{\vec{k}} \a_{\vec{k}'}$ and can be treated following ideas by Kagan and Prokof'ev \cite{Kagan1989}. In the rest of this paper, however, we will discard such terms and assume that the MF polaron state provides a valid starting point to calculate corrections to the adiabatic approximation. Note that the time-dependent ansatz \eqref{eq:DMFvarState} used for our calculations of non-equilibrium dynamics includes corrections due to the additional terms of order $\mathcal{O}(J^* \a^2)$. As a side remark we also mention that from Eq.\eqref{eq:HflucMF} it becomes apparent why, in the absence of driving, $\Omega_{\vec{k}}$ describes the renormalized phonon dispersion in the polaron frame.

\subsubsection{Results: analytical current-force relation}
In the following we will employ Fermi's golden rule to calculate non-adiabatic corrections, corresponding to phonon excitations due to the terms in the second line of Eq.\eqref{eq:HflucMF}. To leading order in $J^*$ we will derive (in \ref{subsubsec:derivation}) the following expression for the current-force relation,
\begin{equation}
v_\text{d}(F) = S_{d-2} 8 \pi  \frac{J_0^{*2}}{aF^2}  \frac{ k^{d-1} V_k^2}{(\partial_k \omega_k)} \l 1 - \text{sinc} (a k)\r + \mathcal{O}(J_0^*)^3, 
\label{eq:gmaPh}
\end{equation}
where $k$ is determined by the condition that $\omega_k = \omega_\text{B}$. Here $J_0^* := \lim_{J \rightarrow 0} J^*(q)$ is the renormalized polaron hopping in the heavy impurity limit (which is independent of $q$), and $S_n=(n+1) \pi^{(n+1)/2} / \Gamma(n/2+3/2)$ denotes the surface area of an $n$-dimensional unit sphere. $\text{sinc}(x)$ is a shorthand notation for the function $\sin(x) / x$.

Importantly, our model yields the closed expression \eqref{eq:gmaPh} for the current-force relation, at least for heavy polarons. Although this limit has been considered before \cite{Bruderer2008}, we are not aware of any such expression describing incoherent polaron transport and derived from first principles. Our result is semi-analytic, in the sense that the prefactor $J_0^*$ has to be calculated numerically from an integral, see Eq.\eqref{eq:CMF0int} below.

In FIG.\ref{fig:vd_F} we compare our numerical results to the semi-analytical expression \eqref{eq:gmaPh}. We obtain excellent agreement for both cases of small and intermediate hopping $J=0.1 c/a$ and $J=0.3 c/a$. For large $J=0.5 c/a$ very close to the subsonic to supersonic transition, larger deviations are found in the weak-driving limit $a F \lesssim c/ \xi$, which in view of the fact that our result Eq.\eqref{eq:gmaPh} is perturbative in the hopping strength $J^*$, does not surprise us. Interestingly, for large force $a F \gtrsim c/\xi$, our semi-analytical theory yields good agreement for all hopping strengths. We will further elaborate on the conditions under which our model works in Subsection \ref{subsubsec:resAndDisc}.

\begin{figure}[t]
\centering
\includegraphics[width=0.5\textwidth]{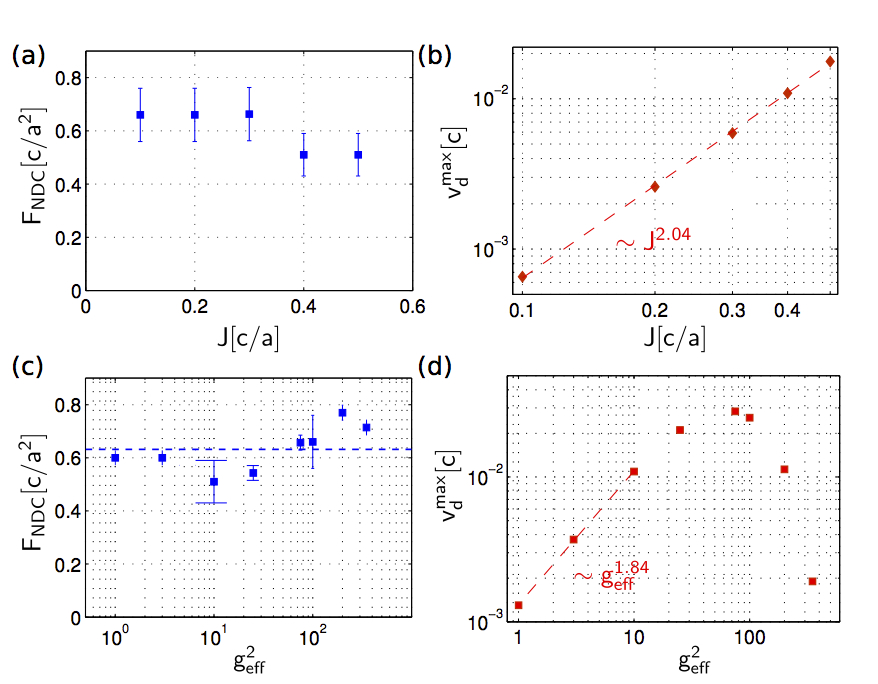}
\caption{(Color online) Dependence of the negative differential conductance peak position, characterized by $F_\NDC$ and $v_\text{d}^{\text{max}}$, on the system parameters; In (a) and (b) the hopping $J$ is varied while the coupling $g_\eff=3.16$ is fixed. In (c) and (d), in contrast, the interaction strength $g_\eff$ is varied while keeping $J=0.4 c/a$ fixed. In (b) and (d) a double-logarithmic scale is used, allowing us to read off the indicated power-law dependencies from best fits to the data (dashed lines), $v_\text{d}^\text{max} \sim J^2$ and $v_\text{d}^\text{max} \sim g_\eff^2$ (for small $g_\eff$). The position $F_\NDC$, in contrast, is only weakly $J$-dependent (a) and we can not observe any clear interaction dependence in (c). The dashed horizontal line in (c) shows the mean of our data. The indicated error bars in (a), (c) are due to the finite mesh used for sampling the underlying current-force relations.}
\label{fig:vdFitRes}
\end{figure}

From Eq.\eqref{eq:gmaPh} we can furthermore obtain a number of algebraic properties of the polaron's current-force relation. To begin with, let us discuss the dependence of the drift velocity on system parameters. Because $V_k \sim g_\eff$ and $J_0^* = J + \mathcal{O}(g_\eff^2)$ we obtain 
\begin{equation}
v_\text{d} \sim g_\eff^2 + \mathcal{O}(g_\eff^4).
\label{eq:vdgeff}
\end{equation}
Moreover, the leading order contribution in the hopping strength scales like
\begin{equation}
v_\text{d} \sim J^2 + \mathcal{O}(J^3).
\label{eq:vdJsq}
\end{equation}
In FIG.\ref{fig:vdFitRes} we investigate the position of the negative differential conductance peak, obtained from the full time-dependent variational simulations of the system. For small hopping $J$ and weak interactions $g_\eff$ we identify power-laws whose exponents agree well with our expectations \eqref{eq:vdgeff}, \eqref{eq:vdJsq} derived above. 

%%%%%%%%%%%%%%%%%%%%%%%%%%%%%%%%%%%%%%%%%%%%%%%%%%%%%
%%%%%%%%%%%%%%%%%%%%%%%%%%%%%%%%%%%%%%%%%%%%%%%%%%%%%

Next, we investigate the behavior in the weak-driving regime. A series expansion of Eq.\eqref{eq:gmaPh} around $F=0$ yields
\begin{equation}
v_\text{d} = F^d g_{\eff}^2 \l J^* \r^2 \xi^2  \frac{a^{3+d} S_{d-2}}{c^{d+1} 6 \sqrt{2} \pi^2} + \mathcal{O}(F^{d+1},J_0^{* 3}).
\label{eq:expdF0anaMod}
\end{equation}
This explains the strong sub-Ohmic behavior we found in Subsection \ref{subsec:numRes}, and furthermore shows that the latter strongly depends on the dimensionality $d$ of the condensate. In particular, for $d=1$, we arrive at Ohmic behavior as found in \cite{Bruderer2008,Bruderer2010,Johnson2011}. The numerical results for $J \leq 0.3 c/a$ in FIG.\ref{fig:vd_F} are also consistent with the power-law $v_\text{d} \sim F^3$ predicted in Eq. \eqref{eq:expdF0anaMod}. Note however that for larger $J$ a comparison of the exponents is difficult because, even for the smallest numerically achievable driving $F$, some residual curvature is left and, more importantly, higher orders in $J^*$ can not simply be neglected.

For large driving, on the other hand, we arrive at the following asymptotic behavior in the continuum limit $\lho = 0$ of the impurity lattice,
\begin{multline}
v_\text{d} = \frac{2^{d/4-1}}{\pi^2} S_{d-2} \l \frac{a}{c} \r^{d/2-2} \xi^{1-d/2} g_\eff^2 \l J_0^{*} \r^2 F^{-3+d/2} \\ 
+ \mathcal{O}(F^{-4+d/2},J_0^{* 3}).
\label{eq:expdFinfanaMod}
\end{multline}
We can not compare our results in FIG.\ref{fig:vd_F} to this power-law, because non-vanishing $\lho \neq 0$ was considered there. Interestingly from a theoretical perspective, as a consequence of Eq.\eqref{eq:expdFinfanaMod}, in $d \geq 6$ dimensions we expect the negative differential conductance peak to disappear. For non-vanishing $\lho$ it reappears of course, but its position may be located at very large $F$. This effect, however, is simply connected to the absence of interacting phonons at the Bloch frequency. Therefore in more than six spatial dimensions coherent Bloch oscillations can never overcome incoherent scattering, in contrast to what we find in lower-dimensional systems.

In the following (\ref{subsubsec:derivation}) we will derive Eq.\eqref{eq:gmaPh}, before we discuss its range of validity as well as possible extensions (in \ref{subsubsec:resAndDisc}).

\subsubsection{Derivation of the current-force relation}
\label{subsubsec:derivation}
To derive Eq.\eqref{eq:gmaPh}, we start by noting that the driving term in Eq.\eqref{eq:HflucMF}, i.e. $F \l \partial_q \alpha^\MF_{\vec{k}}(q(t)) \r$, is $T_\text{B} =2 \pi / \omega_\text{B}$ periodic in time. We can thus expand it in a discrete Fourier-series,
\begin{equation}
 \partial_q \alpha_{\vec{k}}^\MF(q(t)) = \sum_{m=-\infty}^\infty A_{\vec{k}}^{(m)} e^{i \omega_\text{B} m t},
\end{equation}
where the Fourier coefficients read
\begin{equation}
A_{\vec{k}}^{(m)} = \frac{a}{2 \pi} \int_{- \pi /a}^{\pi / a} dq ~ \l \partial_q \alpha_{\vec{k}}^\MF(q) \r e^{i a m q}.
  \label{eq:FourierDriving}
\end{equation}
Using partial integration and a series expansion of $\alpha^\MF_{\vec{k}}$ in $J^*$, we find for $m \geq 0$
\begin{equation}
A_{\vec{k}}^{(m)} = i \delta_{m,1} a J^*_0 \frac{V_k}{\omega_k^2} \l e^{i k_x a} - 1 \r + \mathcal{O}(J^*)^2.
\label{eq:Akexpansion}
\end{equation}
Here we employed that $C^\MF(q) = C^\MF_0 + \mathcal{O}(J^*)$ and $S^\MF(q) = \mathcal{O}(J^*)$ and we used $J_0^* = J e^{-C_0^\MF}$, where
\begin{equation}
C^\MF_0 = \int d^3k~ \frac{V_k^2}{\omega_k^2} \l 1 - \cos ( a k_x ) \r.
\label{eq:CMF0int}
\end{equation}
The coefficients for $m <0$ can be obtained from symmetry, $A_{\vec{k}}^{(-m)} = A_{\vec{k}}^{(m) *} $.

Next, we want to apply Fermi's golden rule to calculate phonon emission due to the driving term $\sim F \l \partial_q \alpha^\MF_{\vec{k}}(q(t)) \r$ in Eq.\eqref{eq:HflucMF}. Before doing so, we notice that the renormalized phonon frequency $\Omega_{\vec{k}}(q(t))$ has a time-dependent contribution. However, we can treat the latter as a perturbation itself and find that to leading order in time-dependent perturbation theory (from which Fermi's golden rule is obtained), it has a vanishing matrix element, $\bra{0} \ad_{\vec{k}} \a_{\vec{k}} \ket{0} = 0$. Then, from Fermi's golden rule we obtain
\begin{equation}
\gamma_\ph = \sum_{m=1}^\infty 2 \pi F^2 \int d^d \vec{k}~ \left| A_{\vec{k}}^{(m)} \right|^2 \delta \l \omega_{k} - m \omega_{\text{B}} \r.
\label{eq:FermiGoldenRule}
\end{equation}
Plugging in Eq.\eqref{eq:Akexpansion} yields our result Eq. \eqref{eq:gmaPh} if we make use of the fact that (to the considered order) phonons are emitted only on the fundamental frequency $\omega_\text{B}$, and using Eq.\eqref{eq:vdFstHarm}, $v_\text{d}=a \gamma_\ph$. In Appendix \ref{apdx:polaronCurrentEAD} a somewhat simpler derivation is presented, which, however, only works in the weakly interacting regime where $J^* = J$ and provided that $F$ is sufficiently small.

\subsubsection{Discussion and extensions}
\label{subsubsec:resAndDisc}
In this paragraph we will further discuss under which conditions our analytical result \eqref{eq:gmaPh} is valid. In particular, we try to understand FIG.\ref{fig:vd_F} (b) in more detail. To this end we suggest an extension of our model, beyond the expression \eqref{eq:FermiGoldenRule} obtained from Fermi's golden rule.

To begin with, we investigate the effect of higher order contributions in the polaron hopping $J^*$. While an analytical series expansion is cumbersome, we note that the truncated Hamiltonian \eqref{eq:HflucMF}, from which we started, is integrable. Since it does not couple different phonon momenta $\vec{k} \neq \vec{k}'$, we only have to solve dynamics of a driven harmonic oscillator at each $\vec{k}$. This can be done numerically using coherent phonon states, and takes into account all orders in the renormalized hopping $J^*$. Compared to a solution of the full time-dependent MF dynamics, which includs couplings between different momenta, it is still cheaper numerically.

In FIG.\ref{fig:vd_F} (b) we also compare our results to such a full solution of the truncated Hamiltonian \eqref{eq:HflucMF} (dashed lines). While for the smallest hopping $J=0.1 c/a$ only small corrections to the result \eqref{eq:gmaPh} from Fermi's golden rule are obtained, we find large corrections for $J=0.3 c/a$ and $J=0.5 c/a$ in weak driving regime $a F \leq c/\xi$ (deviations by up to two orders of magnitude are observed). 

\begin{figure}[t]
\centering
\includegraphics[width=0.5\textwidth]{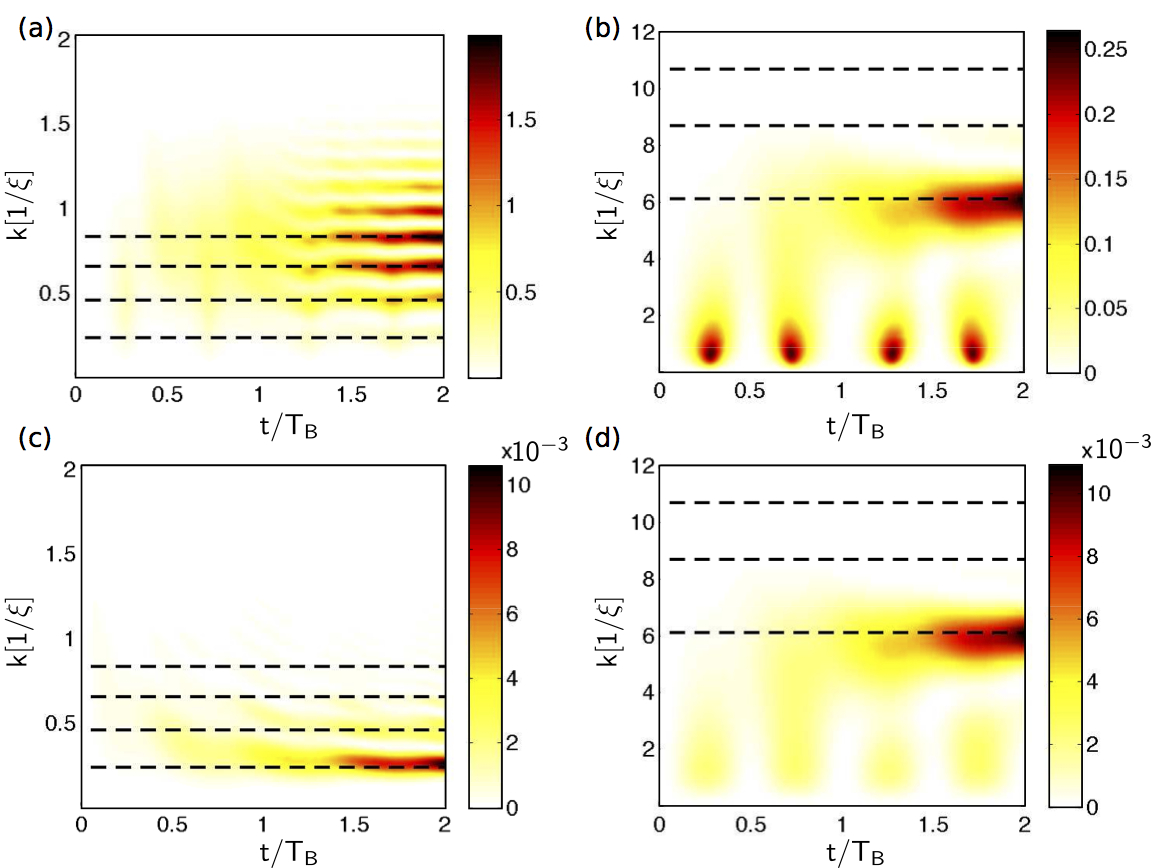}
\caption{(Color online) Phonon energy density $\epsilon(k,t)$ in units of $c$ of the truncated Hamiltonian \eqref{eq:HflucMF} as a function of time and radial momentum $k=|\vec{k}|$. We integrated over the entire momentum shell of radius $k$ and included the measure in the density, i.e. the total phonon energy is $E_\ph(t) = \int dk ~ \epsilon(k,t)$. The results were obtained by solving full dynamics of the truncated Hamiltonian Eq.\eqref{eq:HflucMF} and starting from vacuum. Parameters are $F=0.048 c/a^2$ and $J=0.5 c/a$ in (a), $F=5.4 c/a^2$ and $J=0.5 c/a$ in (b), $F=0.048 c/a^2$ and $J=0.1 c/a$ in (c) and $F=5.4 c/a^2$ and $J=0.1 c/a$ in (d). Positions of the first four resonances $\omega_k = n \omega_\text{B}$ for $n=1,2,3,4$ are indicated by dashed horizontal lines. Other parameters are $g_\eff=3.16$, $\xi=5 a$ and $\lho=a/2$ in all cases.}
\label{fig:compEngyDnsty}
\end{figure}

To understand why this is the case, we first recall that to leading order (i.e. $v_\text{d} \sim J_0^{*2}$) only phonon emission on the fundamental frequency $\omega_\text{B}$ contributes, see Eq.\eqref{eq:Akexpansion}. A higher order series expansion moreover shows that to third order in $J_0^*$, only phonons with frequencies $\omega_k = 2 \omega_\text{B}$ on the second harmonic contribute to $v_\text{d}$. Therefore we expect higher order contributions in $J_0^*$ to lead to phonon emission on higher harmonics. In FIG.\ref{fig:compEngyDnsty} we plot the energy density of emitted phonons, calculated from the truncated Hamiltonian \eqref{eq:HflucMF}. Indeed, for large hopping $J=0.5 c/a$ and weak driving $F=0.048 c/a^2$ we observe multiple resonances in FIG.\ref{fig:compEngyDnsty} (a). For the same force but smaller hopping $J=0.1 c/a$ in contrast, only the fundamental frequency is relevant, see FIG.\ref{fig:compEngyDnsty} (c).

%%%%%%%%%%%%%%%%%%%%%%%%%%%%%%%%%%%%%%%%%%%%%%%%%%%%%
%%%%%%%%%%%%%%%%%%%%%%%%%%%%%%%%%%%%%%%%%%%%%%%%%%%%%

From the comparison in FIG.\ref{fig:vd_F}, we moreover observe that the result Eq.\eqref{eq:gmaPh} from Fermis golden rule, which is perturbative in $J^*$, works surprisingly well in the strong driving regime ($a F \gtrsim c/\xi$), even for hoppings as large as $J=0.5 c/a$ close to the transition to the supersonic regime. To understand why this is the case, we analyze the energy density of phonons for large force $F=5.4 c/a^2$ in FIG. \ref{fig:compEngyDnsty} (b) and (d). We find that in both cases of large and small hopping, $J=0.5 c/a$ in (b) and $J=0.1c/a$ in (d), only emission on the fundamental frequency contributes. This is generally expected in the strong driving regime $a F > c\xi$, as can be seen from a simple scaling analysis. Using Eq.\eqref{eq:FermiGoldenRule} we expect the rate of change of the energy density $\epsilon(k,t)$ for driving with fixed frequency $\omega_\text{B}$ (in $d=3$ dimensions) to scale like
\begin{equation}
\frac{\partial}{\partial t} \epsilon(k,t) \sim k^2 |A_{\vec{k}}^{(m)}|^2 \frac{1}{\partial_k \omega_k}.
\end{equation}
Estimating $A_{\vec{k}}^{(m)} \sim \partial_q \alpha_{\vec{k}}^\MF(q) \sim V_k/\omega_k$ we find the following scalings with momentum,
\begin{equation}
\frac{\partial}{\partial t}  \epsilon(k,t) \sim 
\begin{cases} k &\mbox{if } k \ll 1/\xi , \\ 
\frac{1}{k^3} &\mbox{if } k \gg 1/\xi .
\end{cases} 
\end{equation}
Thus for $\omega_\text{B} > c / \xi$, i.e. for $k > 1 / \xi$, phonon emission on higher harmonics $\omega_k = n \omega_\text{B}$ with $n \geq 2$ is highly suppressed.

 Finally, emission on the fundamental frequency $\omega_k = \omega_\text{B}$ is captured by Fermi's golden rule \eqref{eq:gmaPh} up to corrections of order $J_0^{*6}$, as can be shown using a series expansion of $A_{\vec{k}}^{(m)}$ to second order in $J_0^*$. Thus in the strong driving regime, where mostly the fundamental frequency contributes, only weak $J$-dependence can be expected. This is fully consistent with FIG.\ref{fig:vdFitRes} (d) showing how the negative differential conductance peak varies with $J$. Hardly any deviations from the power-law Eq.\eqref{eq:vdJsq} derived from Fermi's golden rule can be observed there.

\subsection{Insufficiencies of the phenomenological Esaki-Tsu model}
In this subsection we discuss the relation of our results to the phenomenological Esaki-Tsu model \cite{Esaki1970}. While the latter explains some of the qualitative features of the observed current-force relations, we find that it is insufficient for their detailed understanding. Nevertheless, a comparison to this model clarifies how an impurity atom in an optical lattice immersed in a \emph{thermal bath} \cite{Ott2004} differs from a particle immersed in a \emph{superfluid}, as discussed in this paper. In the former case, the Esaki-Tsu relation is valid \cite{Ott2004} and can even be rigorously derived from microscopic models \cite{Ponomarev2006,Kolovsky2008}.

We begin by a brief review of the Esaki-Tsu model and derive its basic predictions for the polaron case. Afterwards we compare these expectations to our numerical results and discuss the differences.

\subsubsection{Phenomenological Esaki-Tsu model}
Esaki and Tsu considered an electron in a periodic lattice, subject to a constant electric field. Using nearest-neighbor tight-binding approximation, the dispersion relation reads $\omega_q = - 2 J \cos \l q a \r$. Because of the external field the particle undergoes Bloch oscillation, so long as incoherent scattering is absent. To include decoherence mechanisms with a rate $1/\tau$, the relaxation time approximation is employed and the following closed expression for the resulting drift velocity was derived \cite{Esaki1970},
\begin{equation}
v_\text{d} = 2 J a \frac{\omega_\text{B} \tau}{1 + \l \omega_\text{B} \tau \r^2}.
\label{eq:EsakiTsu}
\end{equation}

We will not re-derive this result here, however it is instructive to consider the limiting cases $F \rightarrow 0, \infty$. The essence of the relaxation time approximation is the assumption that a wavepacket evolves coherently for a time $\tau$. Then, incoherent scattering takes place and instantly the particle equilibrates in the state of minimal energy, i.e. at $q=0$. In the mean-time the distance traveled in real-space is
\begin{equation}
\Delta x = \int_0^\tau dt ~ \partial_q \omega_q = \frac{2 J}{F} \l 1 - \cos(\omega_{\text{B}} \tau)\r.
\end{equation}
In the weak-driving limit $\omega_{\text{B}} \ll 1/\tau$ we can expand the cosine and find $v_\text{d} = \Delta x / \tau = J \tau a^2 F$, which explains the Ohmic behavior in Eq.\eqref{eq:EsakiTsu}. In the strong driving limit $\omega_{\text{B}} \gg 1/\tau$ on the other hand, we can average out the coherent part of the evolution and set $\cos (\omega_{\text{B}} \tau) \approx 0$. Then we obtain $v_\text{d} = \Delta x / \tau =2 J /(F\tau)$, which captures the large-force limit in Eq.\eqref{eq:EsakiTsu}.

Now we can naively adapt the Esaki-Tsu model to the polaron case, without specifying the origin of the relaxation mechanism. It makes the following predictions for the current-force relation. 
\begin{itemize}
\item[(i)] For weak driving $F \rightarrow 0$, Ohmic behavior $v_\text{d} \sim F$ is expected.
\item[(ii)] For strong driving, negative differential conductance $v_\text{d} \sim 1/F$ is predicted.
\item[(iii)] For intermediate force, a negative differential conductance peak appears, where $d v_\text{d} / d F= 0$.
\item[(iv)] The polaron drift should depend linearly on the effective hopping strength $v_\text{d} \sim J^*$, at least for small hopping $J^* \rightarrow 0$ (for larger hopping, $\tau$ might include $J^*$-dependent corrections). 
\end{itemize}
In the following we will investigate our numerical results more carefully, and show that many of them are not consistent with the simple Esaki-Tsu model, despite the fact that this model has been applied in numerous polaron models before \cite{Bruderer2008,Bruderer2010,Johnson2011}. However, all these points are correctly described by our analytical model of the polaron current.

%%%%%%%%%%%%%%%%%%%%%%%%%%%%%%%%%%%%%%%%%%%%%%%%%%%%%
\begin{figure}[t]
\centering
\includegraphics[width=0.43\textwidth]{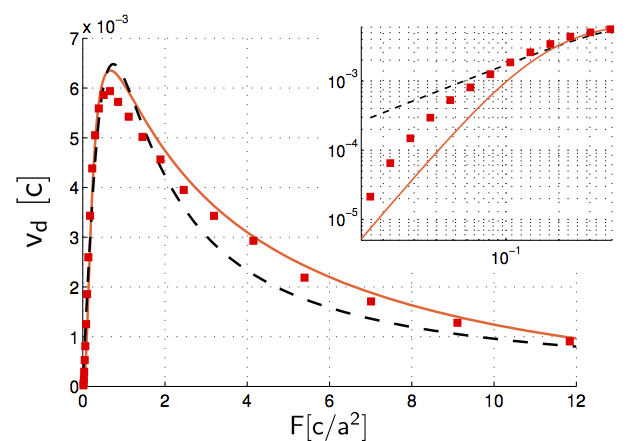}
\caption{(Color online) Best fit of the Esaki-Tsu relation (dashed black line) to our numerically obtained current-force relation $v_\text{d}(F)$ (red squares), where both $J$ and $\tau$ were treated as free parameters in Eq.\eqref{eq:EsakiTsu}. We also show our analytical result Eq.\eqref{eq:gmaPh} (solid orange line), which was obtained from first principles and without any free fitting parameters. While the Esaki-Tsu model can reproduce the negative differential conductance peak, it fits less well in the strong-driving regime. In the inset the same data is shown, but using a double-logarithmic scale. Here the complete failure of the Esaki-Tsu model in the weak-driving limit becomes apparent.
The parameters are $J=0.3 c/a$, $g_\eff=3.16$, $\xi=5 a$ and $\lho=a/\sqrt{2}$ and from the best-fit we obtain $\tau = 1.34 a/c$ and $J |_{\text{fit}} = 0.0065 c/a$.}
\label{fig:compEsakiTsu}
\end{figure}
%%%%%%%%%%%%%%%%%%%%%%%%%%%%%%%%%%%%%%%%%%%%%%%%%%%%%

\subsubsection{Comparison to numerics}
As discussed in Subsection \ref{subsec:numRes} the Esaki-Tsu relation correctly predicts (ii) the existence of negative differential conductance and (iii) a corresponding peak at which $v_\text{d}$ takes its maximum value. This is a direct manifestation of the interplay between coherent transport, which dominates for large $F$, and its incoherent counterpart responsible for the weak-driving behavior. However, we also pointed out already that (i) is inconsistent with the sub-Ohmic behavior observed in our numerics.

In FIG.\ref{fig:compEsakiTsu} we fitted Eq.\eqref{eq:EsakiTsu} to the results of our full solution of the semiclassical dynamical equations \eqref{eq:DMF_EOM}, \eqref{eq:DMF_EOM_phases}. While for moderate driving $F \gtrsim c/a^2$ the shape of the curve can be reproduced by the fit, the comparison for small force (in the inset of FIG.\ref{fig:compEsakiTsu}) clearly shows that the Esaki-Tsu relation can not capture the weak driving regime. Importantly, to get reasonable quantitative agreement, one should treat not only the relaxation time $\tau$, but also the hopping strength $J$ as a free parameter \cite{Bruderer2008}. The resulting best fit $J |_{\text{fit}}$ always yields effective hoppings exceeding the renormalized polaron hopping $J^*=J e^{-C^\MF}$. For instance, in the case shown in FIG.\ref{fig:compEsakiTsu} ($g_{\eff}=3.16$, $J=0.3 c/a$) we find from fitting $J |_{\text{fit}}/J=0.022$ whereas $J^*/J \approx 0.96$ is almost two orders of magnitude larger. Therefore, on a quantitative level, the Esaki-Tsu model completely fails here.

To get a better understanding why the quantitative result from the Esaki-Tsu relation is so far off, we now investigate in detail how the current-force relation $v_{\text{d}}(F)$ depends on our system parameters $g_{\eff}$ and $J$. To this end we consider the position of the negative differential conductance peak, which is characterized by $F_\NDC$ and $v_\text{d}^\text{max}$. From the Esaki-Tsu relation \eqref{eq:EsakiTsu} we would expect $F_\NDC=1/\tau a$ and $v_\text{d}^\text{max}=J^* a$, see (iv).

In FIG. \ref{fig:vdFitRes} (a), (b) we show how $F_\NDC$ and $v_\text{d}^\text{max}$ depend on the hopping strength $J$. While the effect on $F_\NDC$ is rather weak, a power-law very close to $v_\text{d}^\text{max} \sim J^2$ is observed in (b). This is in contradiction to the Esaki-Tsu model, which suggests $v_\text{d}^\text{max} \sim J^*$, since to leading order $J^* \sim J$. It shows that not only $\tau$, but also $J$ should be considered as a fitting parameter in order to describe the numerical curves by the Esaki-Tsu relation \eqref{eq:EsakiTsu}. Physically, however, it is not clear why $J$ should be a free parameter in this equation. Meanwhile, from our analytical model we obtain the correct power-law $v_\text{d} \sim J^2$ for small $J$, see Eq.\eqref{eq:vdJsq}.

In FIG. \ref{fig:vdFitRes} (c), (d) we show the dependence of the negative differential conductance peak on the interaction strength. While no dependence of $F_\NDC$ can be identified (c), we obtain a power-law $v_\text{d}^\text{max} \sim g_{\text{eff}}^2$ for sufficiently weak interactions. From Esaki-Tsu in contrast, we would expect a \emph{decrease} of the polaron drift with the interaction strength, because the latter suppresses the polaron hopping $J^*$. Again, our analytical model can explain the observed power-law, see Eq.\eqref{eq:vdgeff}. It also predicts $v_\text{d}\sim J_0^{*2}$, such that we do indeed expect to find indications of the polaronic dressing for sufficiently large interaction strength. This effect can be observed in (c), where for large $g_\eff$ the incoherent polaron current reaches a maximum value before it becomes strongly suppressed by interactions.

Thus, we have seen that on a quantitative level the Esaki-Tsu model is insufficient for understanding the incoherent polaron current. We attribute the reasonable fit to our data in the moderate driving regime simply to the fact that the Esaki-Tsu relation works on a qualitative level, in the sense that it predicts a negative differential conductance peak.

\section{Summary}
\label{sec:Summary}
In summary we investigated polarons, i.e. mobile impurities dressed by phonons, confined to 1D optical lattices and immersed in a $d$-dimensional BEC. In particular we considered Bloch oscillations of these quasiparticles, which can be observed when a constant force is applied to the impurity. We showed (using an adiabatic approximation) that real-space trajectories of polaron wavepackets provide a tool to measure the renormalized polaron dispersion. By means of a variational MF ansatz we pointed out that the latter is strongly modified at the BZ edges for large impurity-phonon interactions and close to the subsonic to supersonic transition of the polaron. 

Driven by the external force, the phonon cloud has to adjust to the new Bloch wavefunction of the polaron. Since it can not follow its lowest-energy eigenstate completely adiabatically, phonons are emitted. This effect leads to a drift of the polaron along the applied force, and we investigated its dependence on the strength of the driving in detail. In particular, we derived a closed semi-analytical expression for the incoherent polaron current by expanding around the MF polaron solution and employing Fermi's golden rule. A comparison to full time-dependent MF dynamics yields good agreement. From our findings we conclude that the phenomenological Esaki-Tsu model is insufficient for a detailed understanding of the current-force relation, and we pointed out that it completely fails in the weak-driving regime. There, for condensates of dimensionality $d > 1$, we find sub-Ohmic behavior instead of the Ohmic prediction by Esaki and Tsu.

\section*{Acknowledgements}
The authors would like to thank I.Bloch, D. Chowdhury, M. Fleischhauer, R. Schmidt, Y. Shchadilova, A.Widera and S. Will for stimulating discussions. F.G. wants to thank the Harvard University Department of Physics for hospitality during his visit and his PhD supervisor M. Fleischhauer for supporting this research. F.G. is a recipient of a fellowship through the Excellence Initiative (DFG/GSC 266) and gratefully acknowledges financial support from the "Marion K\"oser Stiftung". The authors acknowledge support from the NSF grant DMR-1308435, Harvard-MIT CUA, AFOSR New Quantum Phases of Matter MURI, the ARO-MURI on Atomtronics, ARO MURI Quism program.

\appendix

\section{Derivation of model parameters characterizing impurity-boson interactions}
\label{appdx:InteractionStrengthAndScatteringLength}
In the main text we mentioned that both the scattering length $a_\IB^\eff$ and the effective range $r_\IB^\eff$ of the impurity-boson interaction are modified due to lattice effects, see \cite{Olshanii1998,Massignan2006,Nishida2010}. In the following we discuss in detail how our model parameters $g_\IB$ and $\lho$ (entering $\H_\IB$ implicitly via the Wannier function $w(\vec{r})$ in Eq.\eqref{eq:HIBtwoBody}), which characterize the impurity-boson interaction within our simplified model Eq.\eqref{eq:HIBtwoBody}, relate to the two universal numbers $a_\IB^\eff$ and $r_\IB^\eff$. Since both $a_\IB^\eff$ and $r_\IB^\eff$ can be accessed numerically (see e.g.\cite{Massignan2006,Nishida2010}) or experimentally (see e.g. \cite{Lamporesi2010}), this allows us to make quantitative predictions using our model. Our treatment is analogous to that of \cite{Bauer2013}, where a similar discussion can be found.

To understand the connection between effective model parameters, like $g_\IB$, and universal numbers characterizing inter-particle interactions at low energies, like $a_\IB^\eff$, let us first recall the standard procedure when both the impurity and the boson are unconfined \cite{Bloch2008}. For instance, already when writing the microscopic model in Eq.\eqref{eq:microHam}, we replaced the complicated microscopic impurity-boson interaction potential by a much simpler point-like interaction of strength $g_\IB$. The philosophy here is as follows: when two-particle scattering takes place at sufficiently low energies $k \rightarrow 0$, the corresponding scattering amplitude $f_k$ takes a universal form which is characterized by only a hand-full of parameters -- irrespective of all the microscopic details of the underlying interaction. In particular, for the smallest energies only the asymptotic value of $f_k$ matters, defining the ($s$-wave) scattering length $a_s = - \lim_{k \rightarrow 0} f_k$. 

In an effective model describing low-energy physics only, it is sufficient to capture only the $s$-wave scattering correctly. To this end one may replace the microscopic impurity-boson potential by a simplified pseudo potential, characterized by only a single parameter $g_\IB$. Next, one can calculate the scattering amplitude $f_k(g_\IB)$ expected from this pseudo potential, and to be consistent one has to choose $g_\IB$ such that $a_s(g_\IB) = - \lim_{k \rightarrow 0} f_k(g_\IB)$. This is an implicit equation defining the relation between $a_s$ and $g_\IB$.

In the case when one of the partners (in our case the impurity) is confined to a local oscillator state (a tight-binding Wannier orbital), two body-scattering can be substantially modified. For example, the possibility of forming molecules bound to the local trapping potential gives rise to confinement induced resonances with diverging scattering length, quite similar to Feshbach resonances \cite{Olshanii1998,Massignan2006,Nishida2010}. Importantly for us, this case can be treated in complete analogy to the scenario of free particles described above. The scattering amplitude in the low energy limit is universally given by \cite{Massignan2006}
\begin{equation}
f_k = - \left[ 1 / a_\IB^\eff + i k - r_\IB^\eff k^2 / 2 + \mathcal{O}(k^3) \right]^{-1},
\label{eq:universalFk}
\end{equation}
where $a_\IB^\eff$ denotes the $s$-wave scattering length and $r_\IB^\eff$ is the effective range of the interaction. These two parameters can be calculated from the scattering lengths of unconfined particles, as shown by Massignan and Castin \cite{Massignan2006}, which however requires a full numerical treatment of the two-body scattering problem. Doing so, these authors showed in particular that by varying the lattice depth $V_0$, both parameters can be externally tuned.

Now, instead of going through the complicated microscopic calculations, we introduce a simplified pseudo potential. Motivated by our derivation in the main text, we chose the impurity-boson interaction from Eq.\eqref{eq:HIBtwoBody}. It can be characterized by two parameters, firstly the interaction strength $g_\IB$, and secondly the extent $\lho$ of the involved Wannier functions. In the following both will be determined in such a way that the universal scattering amplitude Eq.\eqref{eq:universalFk} is correctly reproduced. To this end we calculate the latter analytically in Born-approximation and obtain
\begin{equation}
f_k = - \frac{m_\text{B} g_\IB}{2 \pi} \l 1 - k^2 \lho^2/2\r + \mathcal{O}(k^3,g_\IB^2).
\label{eq:fkPerturbatively}
\end{equation}
Comparing Eq.\eqref{eq:fkPerturbatively} to the universal form Eq.\eqref{eq:universalFk} yields the following relations (valid within Born-approximation),
\begin{equation}
g_\IB = \frac{2 \pi}{m_\text{B}} a_\IB^\eff , \quad \qquad \lho^2 = - r_\IB^\eff a_\IB^\eff,
\label{eq:aIBeffvsgIBBORN}
\end{equation}
which define our model parameters (see also \cite{Bauer2013}). 

To derive Eq.\eqref{eq:fkPerturbatively} we assumed the impurity to be localized on a single Wannier site, giving rise to a potential $V_\IB(\vec{r}) =  g_\IB |w(\vec{r})|^2$ seen by the bosons. This is justified in the tight-binding limit, when the hopping $J$ can be treated as a perturbation after handling the scattering problem. Then solving the Lippmann-Schwinger Equation of the scattering problem for a single boson on $V_\IB(\vec{r})$ (perturbatively to leading order in $g_\IB$) yields our result Eq.\eqref{eq:fkPerturbatively}. In order to assure that in the scattering process no higher state in the micro trap is excited, we require the involved boson momenta $k$ to be sufficiently small \cite{Massignan2006},
\begin{equation}
\frac{k^2}{2 m_\text{B}} \ll \omega_0,
\end{equation}
where $\omega_0$ is the micro-trap frequency. Since the involved boson momenta are limited by $k \lesssim 1 / \lho$ from Eq.\eqref{eq:aIBeffvsgIBBORN} we obtain a condition for the interaction strength,
\begin{equation}
\frac{1}{ |r_\IB^\eff| a_\IB^\eff} \ll 2 m_\text{B} \omega_0.
\end{equation}

A comment is in order about the use of the tight-binding approximation in this context. Firstly, to study also cases with stronger hopping along the lattice, the full scattering problem for this case has to be solved. Extending the calculations of \cite{Massignan2006,Nishida2010} to this case, we expect to obtain the same universal form \eqref{eq:universalFk} of the scattering amplitude $f_k$ in the low-energy limit, with modified values for $a_\IB^\eff$ and $r_\IB^\eff$. Nevertheless, the relation Eq.\eqref{eq:aIBeffvsgIBBORN} can still be used to link the the new parameters to the effective model parameters. Secondly, we note that when we discuss approaching the subsonic to supersonic transition in the main text of the paper, this is not necessarily in contradiction to the tight-binding approximation; In fact, the subsonic to supersonic transition takes place around the critical hopping $J_\text{c} a =c$, which is determined solely by properties of the Bose-system. In concrete cases, whether or not tight-binding results may be used, has to be checked for each system individually.

\section{Effective Hamiltonian}
\label{sec:EffecHam}
In this appendix we give a self-contained derivation of the effective Fr\"ohlich type Hamiltonian Eq.\eqref{eq:Hfund} from the main text. It is similar to the derivations given in \cite{Bruderer2008,Tempere2009,Shashi2014RF}

\subsection{Free phonons}
We start from the microscopic Hamiltonian \eqref{eq:Hfund} from the main text, describing the bosonic field $\hat{\phi}$ and the impurity field $\ps$. In the Bose-condensed phase the order parameter is given by the homogeneous BEC density $n_0$. The Bose field operator can be written as $\hat{\phi}(\vec{r}) = \sqrt{n_0} + \hat{\Phi}(\vec{r})$ where $\hat{\Phi}(\vec{r})$ describes quantum fluctuations around the condensate. We calculate the BEC excitation spectrum using standard Bogoliubov theory and write for quantum fluctuations in momentum space
\begin{equation}
 \hat{\Phi}_{\vec{k}} = u_k  \a_{\vec{k}} + v_{k} \ad_{-{\vec{k}}}.
 \label{eq:expQuantFluct}
\end{equation}
The mode functions $u_k,v_{k}$ are given by
\begin{equation}
 u_k=\frac{1}{\sqrt{2}} \sqrt{\frac{1+(k \xi)^2}{k \xi \sqrt{2+(k\xi)^2}}+1} 
\end{equation}
and 
\begin{equation}
 v_{k}=-\frac{1}{\sqrt{2}} \sqrt{\frac{1+(k \xi)^2}{k \xi \sqrt{2+(k\xi)^2}}-1},
\end{equation}
where we introduced the BEC healing length
\begin{equation}
\xi = \l 2 m_\text{B} g_\text{BB} n_0 \r^{-1/2}.
\end{equation}
The excitation spectrum of the BEC is given by
\begin{equation}
\H_\text{B} =\int d^3 \vec{k} ~ \omega_k \ad_{\vec{k}} \a_{\vec{k}},
\end{equation}
and we have chosen the overall energy scale such that the BEC in the absence of the impurity has energy $E=0$. The phonon frequency is $\omega_k = c k \sqrt{1+ \frac{1}{2}\l \xi k \r^2}$ and the speed of sound reads
\begin{equation}
c = \sqrt{g_\text{BB} n_0 / m_\text{B}}.
\end{equation}
Note that, provided they are sufficiently weak, boson-boson interactions can be parametrized by their $s$-wave scattering length $a_{\text{BB}}$ as \cite{Bloch2008}
\begin{equation}
g_{\text{BB}} = \frac{4 \pi a_{\text{BB}}}{m_\text{B}}.
\end{equation}

\subsection{Free impurity}
The free impurity problem can straightforwardly be solved using nearest-neighbor tight-binding approximation. To this end we expand the impurity operator
\begin{equation}
\ps(\vec{r}) = \sum_j w \l \vec{r} - j a \vec{e}_x \r \c_j
 \label{eq:expWannier}
\end{equation}
in terms of Gaussian tight-binding Wannier functions $w(\vec{r})$ (see Eq.\eqref{eq:WannierFct}) with
\begin{equation}
 |w(\vec{r})|^2 = \l \pi \lho^2 \r^{-3/2} e^{- \vec{r}^2 / \lho^2}.
 \label{eq:WannierApdx}
\end{equation}
This yields the impurity Hamiltonian
\begin{equation}
\H_\I = - J \sum_j \l \cd_{j+1} \cd_j + \hc \r - F \sum_j j a ~ \cd_j \c_j.
\end{equation}

\subsection{Impurity-phonon interactions}
Finally we turn to the impurity-boson interaction, which after using the expansion \eqref{eq:expWannier} is described by Eq.\eqref{eq:HIBtwoBody} from the main text. Replacing also quantum fluctuations around the BEC by phonons, see Eq.\eqref{eq:expQuantFluct}, we obtain
\begin{multline}
 \H_\text{IB} = \int d^3 \vec{k} \sum_j \cd_j \c_j e^{i k_x a j} \l \ad_{\vec{k}} + \a_{-\vec{k}} \r V_{k} + \\
 +  n_0 g_\text{IB} + \H_{J-\ph} + \H_{\ph-\ph}.
 \label{eq:HIB}
\end{multline}
Here $\H_{J-\ph}$ denotes phonon induced hoppings (terms of the form $\cd_{j+n} \c_j \a_{\vec{k}}$ with $n \neq 0$), while $\H_{\ph-\ph}$ stands for two-phonon processes (terms of the form $\cd_j \c_j \a_{\vec{k}} \a_{\vec{k}'}$). The interaction strength in Eq.\eqref{eq:HIB} is determined by the form factors $u_k,v_k$ and the Wannier function $w(\vec{r})$,
\begin{equation}
V_k= \sqrt{\frac{n_0}{(2 \pi)^3}} g_\IB \l \frac{(\xi k)^2}{2 + (\xi k)^2} \r^{1/4} \int d^3 \vec{r} e^{i \vec{k} \cdot \vec{r}} |w(\vec{r})|^2.
\end{equation}
Using the Gaussian Wannier function Eq.\eqref{eq:WannierApdx} from above we obtain
\begin{equation}
 V_k = (2 \pi)^{-3/2} \sqrt{n_0} g_\IB \l \frac{(\xi k)^2}{2 + (\xi k)^2} \r^{1/4} e^{-k^2 \lho^2/4}.
 \label{eq:VkApdx2}
\end{equation}
The relation between $g_\IB$ and a measurable scattering length was discussed in Appendix \ref{appdx:InteractionStrengthAndScatteringLength}.

To obtain the final polaron Hamiltonian Eq.\eqref{eq:Hfund}, we neglect phonon induced tunneling as well as two-phonon processes. In the following two paragraphs we discuss under which conditions this is justified.

\subsubsection{Two-phonon processes}
Neglecting two-phonon processes is justified if the phonon density $n_\text{ph}$ is much smaller than the BEC density, i.e. for
\begin{equation}
n_\text{ph} \ll n_0.
\label{eq:nphNBEC}
\end{equation}
In this case scattering events involving a boson from the condensate dominate over phonon-phonon terms. 

In order to estimate the phonon density due to quantum depletion $\delta N_0$ of the condensate around the impurity, let us calculate the latter perturbatively from the term linear in phonon operators in Eq.\eqref{eq:HIB},
\begin{equation}
\delta N_0 \approx \int d^3 \vec{k} ~ \l \frac{V_k}{\omega_k} \r^2.
\end{equation}
Assuming that the typical length scale associated with the Wannier function $\lho / \sqrt{2} \leq \xi$ is smaller than the healing length we find that $V_k$ saturates at $k \approx 1/\xi$, see \eqref{eq:VkApdx2}, while $\omega_k$ changes from linear $\sim k$ to quadratic $\sim k^2$ behavior. Therefore only momentum modes with $k \lesssim 1/\xi$ contribute substantially to quantum depletion in the vicinity of the impurity. Consequently depletion takes place on a spatial scale set by $\xi$ and we require
\begin{equation}
n_\text{ph} \approx \delta N_0 \xi^{-3} \ll n_0.
\end{equation}
Integrating only up to $1/\xi$ in $k$-space and using the scaling $V_k \sim \sqrt{k}$ and $\omega_k \sim k$ with the correct prefactors, valid for $k \lesssim 1/\xi$, we obtain the estimate
\begin{equation}
|g_\IB| \xi^{-3} \ll 7.5 c / \xi.
\label{eq:condgIB}
\end{equation}
 This condition is similar to the one derived in \cite{Bruderer2007}, $|g_\IB | \xi^{-3} \ll 4 c / \xi$.

\subsubsection{Phonon-induced tunneling}
Phonon-induced tunneling is described by
\begin{equation}
\H_{J-\ph} = \sum_{i > j}  \cd_i \c_j \int d^3 \vec{k} ~ e^{i k_x a j} \l \ad_{\vec{k}} + \a_{-\vec{k}} \r V_{k}^{(i-j)} + \hc,
\end{equation}
where the corresponding scattering amplitudes read
\begin{equation}
V_k^{(n)} =  V_k \frac{\int d^3 \vec{r} ~ w^*(\vec{r}-n a \vec{e}_x) e^{i \vec{k} \cdot \vec{r}} w(\vec{r})}{\int d^3 \vec{r} ~  e^{i \vec{k} \cdot \vec{r}} |w(\vec{r})|^2 }
\label{eq:Vkn}
\end{equation}
for integer $n=...,-1,0,1,...$ ~ . 

We may neglect such phonon-induced tunnelings, if the scattering amplitude $V_k^{(0)}$ dominates over all those involving tunneling, $V_k^{(n)}$ with $n \neq 0$. Their ratio is given by matrix elements of $e^{i \vec{k} \cdot \vec{r}}$ with respect to the Wannier functions and we obtain the condition
\begin{equation}
\frac{|V_k^{(n)}|}{|V_k|} = \frac{ |\bra{w_n} e^{i \vec{k} \cdot \vec{r}} \ket{w_0}| }{| \bra{w_0} e^{i \vec{k} \cdot \vec{r}} \ket{w_0} | } \stackrel{!}{\ll} 1,
\end{equation}
when phonon-induced tunneling can be discarded. In the tight-binding limit, this is usually fulfilled when Wannier functions are well localized.

\section{Static MF polarons}
\label{sec:StatMFpolaron}
In this appendix we derive the MF self-consistency equation \eqref{eq:MFselfCons} from the main text. To this end we have to calculate the variational energy,
\begin{equation}
 \HM[\alpha_{\vec{k}}](q) = \prod_{\vec{k}} \bra{\alpha_{\vec{k}}} \H_q \ket{\alpha_{\vec{k}}}.
\end{equation}
The main obstacle is the treatment of the non-linear term $\sim \cos \ad \a$ in the Hamiltonian $\H_q$ \eqref{eq:polaronHam}, for which we find
\begin{multline}
\prod_k \bra{\alpha_{\vec{k}}} \cos \l a q- a \int d^3\vec{k}'~k_x' \hat{n}_{\vec{k}'} \r \ket{\alpha_{\vec{k}}} \\ = e^{-C[\alpha_{\vec{\kappa}}]} \cos \l  a q - S[\alpha_{\vec{\kappa}}] \r.
\label{eq:cosCoherentExpectation}
\end{multline}
The functionals $C[\alpha_{\vec{\kappa}}]$ and $S[\alpha_{\vec{\kappa}}]$ appearing in this expression were defined in the main text, see Eqs.\eqref{eq:Cqdef}, \eqref{eq:Sqdef}.

To proof the result \eqref{eq:cosCoherentExpectation}, let us first focus on a single mode and replace $\int d^3 \vec{k}' ~ k_x' \ad_{\vec{k}'} \a_{\vec{k}'}$ by $k \ad \a$ for simplicity. Next we write the cosine in terms of exponentials, for which it is then sufficient to show that
\begin{equation}
\bra{\alpha} \exp \left[ i a k \ad \a \right] \ket{\alpha} = \exp \left[ - |\alpha|^2 \l 1 - e^{i a k} \r \right].
\end{equation}
This is most easily achieved by expanding coherent states $\ket{\alpha}$ in the Fock basis $\ket{n}$,
\begin{equation}
\ket{\alpha} = e^{- |\alpha|^2/2} \sum_{n=0}^\infty \frac{\alpha^n}{\sqrt{n!}} \ket{n},
\end{equation}
from which we can read off the relation,
\begin{multline}
\bra{\alpha} e^{i a k \ad \a} \ket{\alpha} = e^{- |\alpha|^2} \sum_{n=0}^\infty \frac{e^{i a k n} |\alpha|^{2n}}{n!} \\
 =  \exp \left[ - |\alpha|^2 \l 1 - e^{i a k} \r \right].
\end{multline}
This result can easily be generalized to the multimode case with $\int d^3 \vec{k} ~k_x  \hat{n}_{\vec{k}}$ appearing in the argument of the cosine, when use is made of the commutativity of phonon modes at different momenta.

Using the result Eq.\eqref{eq:cosCoherentExpectation} we end up with the variational Hamiltonian
\begin{multline}
 \HM[\alpha_{\vec{\kappa}},\alpha_{\vec{\kappa}}^*] = - 2 J e^{-C[\alpha_{\vec{\kappa}}]} \cos \l a q -S[\alpha_{\vec{\kappa}}] \r \\ +\int d^3k ~ \left[ \omega_k |\alpha_k|^2 + V_k \l \alpha_k + \alpha_k^* \r \right].
 \label{eq:HMfunctional}
\end{multline}
The MF self consistency equations can now easily be obtained by demanding vanishing functional derivatives,
\begin{equation}
\frac{\delta  \HM[\alpha_k,\alpha_k^*]}{\delta \alpha_k} = \frac{\delta  \HM[\alpha_k,\alpha_k^*]}{\delta \alpha_k^*} \stackrel{!}{=} 0,
\end{equation}
which readily yields Eq.\eqref{eq:MFselfCons} from the main text,
\begin{equation}
\alpha_{\vec{k}}^\MF = - \frac{V_k}{\Omega_{\vec{k}}[\alpha^\MF_{\vec{\kappa}}]}.
\end{equation}
Plugging this result into the definitions of $C[\alpha_{\vec{\kappa}}]$ and $S[\alpha_{\vec{\kappa}}]$ yields the coupled set of self-consistency equations \eqref{eq:selfConsC},\eqref{eq:selfConsS} for $C^\MF$ and $S^\MF$.

\section{Time dependent variational principle}
\label{sec:DiracVarPrinc}
To derive the equations of motion for the time-dependent variational phonon state Eq.\eqref{eq:DMFvarState}, we apply Dirac's variational principle (see e.g. \cite{,Jackiw1980}). It states that, given a possibly time-dependent Hamiltonian $\H(t)$, the dynamics of a quantum state $\ket{\psi(t)}$ (which can alternatively be described by the Schr\"odinger equation) can be obtained from the variational principle
\begin{equation}
\delta \int dt ~ \bra{\psi(t)} i \partial_t - \H(t) \ket{\psi(t)}=0.
\label{eq:DiracVarPrinciple}
\end{equation}
We reformulate this in terms of a Lagrangian action $\mathcal{L} = \bra{\psi(t)} i \partial_t - \H(t) \ket{\psi(t)}$ and obtain
\begin{equation}
\delta \int dt~ \mathcal{L} = 0.
\end{equation}
When using a variational ansatz $\ket{\psi(t)} = \ket{\psi[x_j(t)]}$ defined by a general set of time-dependent variational parameters $x_j(t)$, we obtain their dynamics from the Euler-Lagrange equations of the classical Lagrangian $\mathcal{L}[x_j,\dot{x}_j,t]$. 

We note that there is a global phase degree of freedom: when $\ket{\psi(t)}$ is a solution of \eqref{eq:DiracVarPrinciple}, then so is $e^{-i \chi(t)} \ket{\psi(t)}$ because the Lagrangian changes as $\mathcal{L} \rightarrow \mathcal{L} + \partial_t \chi(t)$. To determine the dynamics of $\chi(t)$ we note that for the exact solution $\ket{\psi_{\text{ex}}(t')}$ of the Schr\"odinger equation it holds
\begin{equation}
\int_0^t \mathcal{L}(t')=0,
\label{eq:EOMphase_SM}
\end{equation}
for all times $t$, i.e. $\mathcal{L}=0$. This equation can then be used to determine the dynamics of the overall phase for variational states.

Now we can construct the Lagrangian $\mathcal{L}$ for the variational coherent phonon state, Eq.\eqref{eq:DMFvarState} in the main text. Using the following identity for coherent states $\ket{\alpha}$
\begin{equation}
\bra{\alpha} \partial_t \ket{\alpha} = \frac{1}{2} \l \dot{\alpha} \alpha^* - \dot{\alpha}^* \alpha \r,
\end{equation}
we obtain
\begin{multline}
\mathcal{L}[\alpha_{\vec{k}},\alpha_{\vec{k}}^*,\dot{\alpha}_{\vec{k}},\dot{\alpha}_{\vec{k}}^*,t] = \partial_t \chi_q - \HM[\alpha_{\vec{k}},\alpha_{\vec{k}}^*] \\ - \frac{i}{2} \int d^3 \vec{k} \l \dot{\alpha}_{\vec{k}}^*\alpha_{\vec{k}} - \dot{\alpha}_{\vec{k}} \alpha_{\vec{k}}^* \r
\label{eq:LcalcApdx}
\end{multline}
where the Hamiltonian $\HM$ is given by \eqref{eq:HMfunctional}. Using Eq.\eqref{eq:LcalcApdx} the Euler-Lagrange equations yield the equations of motion \eqref{eq:DMF_EOM} from the main text,
\begin{equation}
 i \partial_t \alpha_{\vec{k}}(t) =\Omega_{\vec{k}}[\alpha_{\vec{\kappa}}(t)] ~  \alpha_{\vec{k}}(t) + V_k.
\label{eq:DMF_EOM_SM}
\end{equation}
Moreover, as described above, Eq.\eqref{eq:EOMphase_SM} yields equations of motion for the global phases Eq.\eqref{eq:DMF_EOM_phases} given in the main text,
\begin{equation}
\partial_t \chi_q = \frac{i}{2} \int d^3\vec{k} \l \dot{\alpha}_{\vec{k}}^*\alpha_{\vec{k}} - \dot{\alpha}_{\vec{k}} \alpha_{\vec{k}}^* \r+ \HM[\alpha_{\vec{k}},\alpha_{\vec{k}}^*].
\end{equation}
Using the equations of motion \eqref{eq:DMF_EOM_SM} for $\alpha_{\vec{k}}$, this simplifies somewhat and we obtain
\begin{widetext}
\begin{multline}
\partial_t \chi_q = 2 J e^{-C[\alpha_{\vec{k}}(t)]} \Bigl[ S[\alpha_{\vec{k}}(t)] \sin \l a q - \omega_{\text{B}} t - S[\alpha_{\vec{k}}(t)] \r   - \l 1 + C[\alpha_{\vec{k}}(t)]\r \cos \l a q - \omega_{\text{B}} t - S[\alpha_{\vec{k}}(t)] \r \Bigr]  + \text{Re} \int d^3 \vec{k}~ V_{k} \alpha_{\vec{k}}.
\end{multline}
\end{widetext}

\section{Impurity density}
\label{sec:ImpDensity}
In this appendix we derive Eq.\eqref{eq:njtDMF} from the main text, which allows us to calculate the impurity density from the time-dependent overlaps $A_{q_2,q_1}(t)$. For the definition of the latter, let us recall that we work in the polaron frame throughout, where the quantum state is of the form
\begin{equation}
\ket{\Psi(t)} = \sum_{q \in \BZ} f_q \cd_q \ket{0}_c \otimes \ket{\Psi_q(t)}_a.
\label{eq:PsitPolaronFrame}
\end{equation}
Here $\ket	{0}_c$ denotes the impurity vacuum and $\ket{\Psi_q(t)}_a$ is a pure phonon wavefunction. The corresponding time-dependent overlaps are defined as
\begin{equation}
A_{q_2,q_1}(t) = ~_{a}\langle \Psi_{q_2}(t) \ket{\Psi_{q_1}(t)}_a,
\end{equation}
see also Eq.\eqref{eq:defNonEqGreen}.

In order to calculate the impurity density in the lab frame, $n_j = \langle \cd_j \c_j \rangle$, we have to transform the operator $\cd_j \c_j$ to the polaron frame first. Keeping in mind that we moreover applied the time-dependent unitary transformation $\hat{U}_\text{B}(t)$ Eq.\eqref{eq:BOtransform}, we thus arrive at
\begin{equation}
n_j=\langle \cd_j \c_j \rangle_\text{lab} =  \bra{\Psi(t)} \hat{U}_\text{LLP}^\dagger \hat{U}_\text{B}^\dagger(t) \cd_j \c_j \hat{U}_\text{B}(t) \hat{U}_\text{LLP} \ket{\Psi(t)}.
\end{equation}
We proceed by writing the impurity operators $\c_j$ in their Fourier components, see Eq.\eqref{eq:BOoscillatingOps}, and plug Eq.\eqref{eq:PsitPolaronFrame} into the last expression,
\begin{widetext}
\begin{equation}
n_j(t) = \frac{a}{L} \sum_{q_1,q_2 \in \BZ} e^{-i \l q_1 - q_2 \r a j}  \sum_{q_3,q_4 \in \BZ} f_{q_3}^* f_{q_4} ~ ~_{a}\bra{\Psi_{q_3}(t)}_c\bra{0} ~ \c_{q_3}  \hat{U}_\text{LLP}^\dagger \hat{U}_\text{B}^\dagger(t) ~ \cd_{q_2} \c_{q_1} ~  \hat{U}_\text{B}(t) \hat{U}_\text{LLP} ~ \ket{0}_c \ket{\Psi_{q_4}(t)}_a.
\label{eq:njtResultAppdx}
\end{equation}
\end{widetext}
To simplify this expression, we note that
\begin{equation}
\hat{U}^\dagger_\text{B}(t) \cd_{q_2} \hat{U}_\text{B}(t) = \cd_{q_2+\omega_\text{B} t}
\end{equation}
and analogously for $\c_{q_1}$. Thus, by relabeling indices $q_{1,2} \rightarrow q_{1,2} + \omega_\text{B} t$ in Eq.\eqref{eq:njtResultAppdx}, we can completely eliminate $\hat{U}_\text{B}(t)$ from the equations above. 

To deal with the Lee-Low-Pines transformation, let us introduce an eigen-basis consisting of states $\ket{P}$ where the total phonon momentum is diagonal,
\begin{equation}
\int d^3 \vec{k} ~ k_x \ad_{\vec{k}} \a_{\vec{k}} \ket{P} = P \ket{P}.
\label{eq:PPPP}
\end{equation}
Of course, for each value of $P$ there is a large number of states denoted by $\ket{P}$ with this property \eqref{eq:PPPP}. Importantly, the Lee-Low-Pines transformation Eq.\eqref{eq:polaronTrofo} can now easily be evaluated in this new basis, where
\begin{equation}
\bra{P} \hat{U}_\text{LLP} \ket{P'} = e^{i \hat{X} P} \delta_{P,P'}
\end{equation}
and for simplicity we used a discrete set of phonon modes. We can make use of this result by formally introducing a unity in this basis,
\begin{equation}
\sum_P \ket{P}\bra{P} = \hat{1},
\label{eq:unityPapdx}
\end{equation}
allowing us to write
\begin{multline}
\hat{U}_\text{LLP}^\dagger  \cd_{q_2} \c_{q_1}  \hat{U}_\text{LLP} = \sum_{P,P'} \ket{P} \bra{P} \hat{U}_\text{LLP}^\dagger  \cd_{q_2} \c_{q_1}  \hat{U}_\text{LLP} \ket{P'} \bra{P'} \\
=  \sum_{P} \ket{P} \bra{P} e^{-i P \hat{X}} \cd_{q_2} \c_{q_1} e^{ i P \hat{X}}.
\end{multline}
Next, using $e^{-i P \hat{X}} \c_{q_1} e^{ i P \hat{X}} = \c_{q_1 + P}$, we obtain
\begin{equation}
\hat{U}_\text{LLP}^\dagger  \cd_{q_2} \c_{q_1}  \hat{U}_\text{LLP} = \sum_{P} \ket{P} \bra{P} \cd_{q_2+P} \c_{q_1+P}.
\end{equation}
Using this identity after introducing unities \eqref{eq:unityPapdx} in Eq.\eqref{eq:njtResultAppdx}, we find after relabeling summation indices $q_{1,2} \rightarrow q_{1,2} + P$ that
\begin{multline}
n_j(t) = \frac{a}{L} \sum_{q_1,q_2 \in \BZ} e^{-i \l q_1 - q_2 \r a j}  \sum_{q_3,q_4 \in \BZ} f_{q_3}^* f_{q_4} \\
\times ~_{c}\bra{0} \c_{q_3} \cd_{q_2} \c_{q_1} \cd_{q_4} \ket{0}_c ~ ~_{a}\bra{\Psi_{q_3}(t)} \Psi_{q_4}(t) \rangle_a.
\label{eq:njtResultAppdx2}
\end{multline}
After simplification of the impurity operators we obtain the desired result,
\begin{equation}
n_j(t) =  \frac{a}{L} \sum_{q_1,q_2 \in \BZ} e^{-i \l q_1 - q_2 \r a j}  f_{q_2}^* f_{q_1} A_{q_2,q_1}(t).
\end{equation}

\section{Adiabatic wavepacket dynamics}
\label{sec:AdiabaticWavePacket}
In this appendix we present the detailed calculation leading to the expression for the adiabatic impurity density \eqref{eq:adiabImpDsty} given in the main text. To this end we first calculate the time-dependent overlaps from Eqs.\eqref{eq:cohPart}, \eqref{eq:incohPart},
\begin{equation}
A_{q_2,q_1}(t) = \bra	{\Psi_{q_2}(t)} \Psi_{q_1}(t) \rangle = \mathcal{A}_{q_2,q_1} \mathcal{D}_{q_2,q_1},
\end{equation}
and use Eq.\eqref{eq:njtDMF} together with a suitable initial impurity wavefunction $\psi_j^\text{in}$. 

We start from an impurity wavepacket in the band minimum of the bare impurity, and assume its width (in real space)  $L_\I \gg a$ by far exceeds the lattice spacing $a$. In this case the width in (quasi-) momentum space is $\delta q \approx 2 \pi/L_\I \ll 2 \pi / a$. Moreover we can treat $a j \rightarrow x$ as a continuous variable and write
\begin{equation}
\psi^\text{in}(x)= (2\pi)^{-1/4} L_\I^{-1/2} \exp \l - \frac{x^2}{4 L_\I^2}\r,
\end{equation}
where the following normalization was chosen,
\begin{equation}
\int_{-\infty}^\infty dx~ |\psi^\text{in}(x)|^2 =1.
\end{equation}
By Fourier-transforming the initial impurity wavefunction we obtain the amplitudes
\begin{equation}
f_q = \frac{1}{\sqrt{2 \pi}} \int_{-\infty}^\infty  dx~ e^{i q x} \psi^\text{in}(x),
\end{equation}
such that the impurity density \eqref{eq:njtDMF} becomes
\begin{equation}
n(x,t) =\int_{-\infty}^\infty \frac{dq_2 dq_1}{2 \pi} ~ e^{i \l q_2 - q_1 \r x} f_{q_2}^* f_{q_1} A_{q_2,q_1}(t).
\label{eq:nxtConti}
\end{equation}
Since the width $\delta q \ll 2 \pi / a$ of the wavepacket is much smaller than the size of the BZ, we approximated $\int_\text{BZ} dq \approx \int_{-\infty}^\infty dq$ in this step.

Using the adiabatic wavefunction \eqref{eq:adiabApprx}, the phases of $A_{q_2,q_1}(t)$ read
\begin{equation}
\mathcal{A}_{q_2,q_1}(t) = \exp \left[ i \int_0^t dt' ~ \HMF(q_2(t')) - \HMF(q_1(t')) \right],
\end{equation}
and the amplitude is given by
\begin{equation}
\mathcal{D}_{q_2,q_1}(t)=\exp \left[ - \frac{1}{2} \int d^3 \vec{k} \left\{ \alpha^\MF_{\vec{k}}(q_2(t)) - \alpha^\MF_{\vec{k}}(q_1(t)) \right\}^2 \right].
\end{equation}
Due to the small width $\delta q$ of the polaron wavepacket in quasimomentum space we can expand the expressions in the exponents in powers of the difference $q_2(t)-q_1(t)=q_2-q_1$. Note that $\log \mathcal{A}_{q_2,q_1}$ ($\log \mathcal{D}_{q_2,q_1}$) is antisymmetric (symmetric) under exchange of $q_{2}$ and $q_1$. To second order in $|q_2-q_1|$ we obtain
\begin{flalign}
\mathcal{A}_{q_2,q_1}(t) &= \exp \left[ i \l q_2 - q_1\r \int_0^t dt' ~ \partial_{q} \HMF(q_1(t'))  \right], \\
\mathcal{D}_{q_2,q_1}(t) & = \exp \left[ - \frac{1}{2}  \l q_2 - q_1\r^2 \int d^3 \vec{k} \l \partial_q \alpha^\MF_{\vec{k}}(q_1(t)) \r^2 \right].
\end{flalign}
Since only $q_1\approx q_2 \approx 0$ contributes substantially in $f_{q_2}^*f_{q_1}$ we further approximate 
\begin{equation}
\partial_{q} \HMF(q_1(t')) \approx \partial_{q} \HMF(-F t')
\end{equation}
and analogously in $\partial_q \alpha_{\vec{k}}^\MF$. Thus we obtain
\begin{equation}
A_{q_2,q_1}(t) = \exp \left[ -i \l q_2 - q_1 \r X(t) - \frac{1}{2} \l q_2 - q_1 \r^2 \Gamma^2(t) \right],
\label{eq:Aq2q1ApdxRes}
\end{equation}
with $\Gamma^2(t)$ defined in Eq.\eqref{eq:GtAd} in the main text and
\begin{equation}
X(t) = X(0)-\int_0^t dt'~\left.  \partial_q \HMF \right\vert_{q=-F t'}.
\end{equation}
Evaluating this integral exactly yields the expression for $X(t)$ given in the main text, Eq.\eqref{eq:XtAd}.

Using the last expression for $A_{q_2,q_1}$ \eqref{eq:Aq2q1ApdxRes} to perform momentum integrals $dq_1 dq_2$ in \eqref{eq:nxtConti} finally yields the adiabatic impurity density
\begin{equation}
n(x,t) = e^{ - \frac{(x-X(t))^2}{2 \l L_\I^2 + \Gamma^2(t) \r } } \left[ 2 \pi \l L_\I^2 + \Gamma^2(t) \r \right]^{-1/2}
\end{equation}
as we claimed in the main text.

\section{Alternative derivation of polaron current in weak-coupling and small-hopping limit}
\label{apdx:polaronCurrentEAD}
In this appendix we give an alternative derivation of the analytical current-force relation Eq.\eqref{eq:gmaPh} introduced in the main text. The following treatment is somewhat simpler conceptually, however it is only valid in the limit of small force $F$ and weak interactions $g_\eff \rightarrow 0$. For simplicity we restrict our discussion to $d=3$ dimensions, but all arguments can easily generalized to arbitrary $d$.

The idea is to start from the Hamiltonian in Eq.\eqref{eq:Hfund} in the lab frame, i.e. before applying the Lee-Low-Pines transformation. Then we can consider the limit $g_\eff \rightarrow 0$, where to first approximation the impurity can be treated as being independent of the phonons. If we moreover assume that the particle is sufficiently heavy, i.e. $J$ is small, we may neglect fluctuations of the impurity position and approximate the latter by its mean,
\begin{equation}
x(t) \approx \langle x(t) \rangle = \frac{2 J}{\omega_{\text{B}}} \cos \l \omega_\text{B} t \r.
\end{equation}

To describe the interactions of the impurity with phonons, we now plug the last equation into Eq.\eqref{eq:Hfund} and obtain
\begin{multline}
 \H(t) \approx \int d^3 \vec{k} \Bigl\{ \omega_k \ad_{\vec{k}} \a_{\vec{k}} +  e^{i k_x \langle x(t) \rangle} \l \ad_{\vec{k}} + \a_{-\vec{k}} \r V_{k} \Bigr\}.
  \label{eq:HfundAppdx}
\end{multline}
Then we can expand the exponential in orders of the hopping, $e^{i k_x \langle x(t) \rangle} \approx 1 + i k_x \frac{2 J}{\omega_{\text{B}}} \cos \l \omega_\text{B} t \r + \mathcal{O}(J^2)$, and treat the resulting oscillatory term using Fermi's golden rule. As a result we obtain, using $v_\text{d} = a \gamma_\text{ph}$ as described in the main text,
\begin{equation}
v_\text{d} = \frac{8 \pi^2 a}{3} \frac{J^2}{F^2}  \frac{k^4 V_k^2}{\l \partial_k \omega_k \r}.
\label{eq:vdEAD}
\end{equation}

Now as in the main text, we can perform a series expansion of the resulting expression \eqref{eq:vdEAD} in the driving force $F$. In the weak driving limit we obtain
\begin{equation}
v_\text{d} = \frac{a^6}{c^4 3 \pi \sqrt{2}} g_\eff^2 J^2 \xi^2 F^3 + \mathcal{O}(F^4).
\end{equation}
Notably, this is exactly the same expression as Eq.\eqref{eq:expdF0anaMod} from our calculation in the main text, except that $J$ appears instead of $J_0^*$. 

In the strong driving limit, in contrast, we obtain a different power-law than in the main text Eq.\eqref{eq:expdFinfanaMod},
\begin{equation}
v_\text{d} = \frac{2^{1/4}}{3 \pi} c^{-1/2} a^{5/2} \xi^{-3/2} g_\eff^2 J^2 F^{-1/2} + \mathcal{O}(F^{-3/2}),
\end{equation}
where we used the impurity continuum limit again, i.e. $\lho \rightarrow 0$. The reason why we do not reproduce the result from the (more involved) calculation in the main text is that expanding the exponential below Eq.\eqref{eq:HfundAppdx} contains a small $k$ approximation as well.

\end{document}